\PassOptionsToPackage{hypertexnames=false}{hyperref}
\documentclass[fleqn,usenatbib,useAMS]{mnras}

\usepackage[T1]{fontenc}
\usepackage{newtxtext,newtxmath}
\usepackage{ulem} 
\usepackage{graphicx,color} 
\usepackage{dcolumn} 
\usepackage{bm} 

\def\aap{A\&A}
\def\apjl{ApJ}

\def\apjs{ApJS}

\usepackage{color}
\newcommand{\blue}[1]{#1} 

\def\P#1#2{\dfrac{\partial #1}{\partial #2}}
\def\PP#1#2{\dfrac{\partial^2 #1}{\partial #2^2}}

\title[Orbital Bias of Baroclinic Double-Cell Circulation in Binary Stars]
{Orbital Bias of Baroclinic Double-Cell Circulation in Binary Stars}

\author[N. Yasutake et al.]{
Nobutoshi Yasutake$^{1,2}$\thanks{E-mail: nobutoshi.yasutake@p.chibakoudai.jp},
Shoichi Kitagawa$^{3}$,
Masaki Iwasawa$^{4}$,
Kotaro Fujisawa$^{5}$,
\newauthor
Hirotada Okawa$^{6}$,
Miasa Ogata$^{7}$,
Shoichi Yamada$^{7}$\\
$^{1}$Department of Physics, Chiba Institute of Technology, 2-1-1 Shibazono, Narashino, Chiba 275-0023, Japan\\
$^{2}$Advanced Science Research Center, Japan Atomic Energy Agency, Tokai, Ibaraki 319-1195, Japan\\
$^{3}$Department of Computer Science, Chiba Institute of Technology, Tsudanuma Campus, 2-17-1 Tsudanuma, Narashino, Chiba 275-8588, Japan\\
$^{4}$National Institute of Technology, Matsue College, 214-4 Nishi-Ikuma, Matsue, Shimane 690-8518, Japan\\
$^{5}$Kogakuin University, Hachioji Campus, 2665-1 Nakano, Hachioji, Tokyo 192-0015, Japan\\
$^{6}$Aomori University, Tokyo Campus, 2-10-1 Seishincho, Edogawa, Tokyo 134-0087, Japan\\
$^{7}$Advanced Research Institute for Science and Engineering, Waseda University, Okubo 3-4-1, Shinjuku, Tokyo 169-8555, Japan
}

\date{}
\pubyear{2026}

\begin{document}
\label{firstpage}
\pagerange{\pageref{firstpage}--\pageref{lastpage}}
\maketitle

\begin{abstract}

We study baroclinicity-driven double-cell circulation in self-gravitating equal-mass binaries and examine how orbital rotation biases this circulation. Using smoothed particle hydrodynamics, we obtain quasi-steady states for an imposed entropy profile in each star. The baroclinicity develops persistent large-scale two-cell circulation in the stellar interiors. The circulation pattern is nearly symmetric with respect to reflection across the orbital plane. Within the orbital plane, however, the circulation pattern lacks reflection symmetry, although the double-cell circulation persists. This asymmetry arises from the orbital motion of the binary stars. A minimal linear model reproduces the direction of the skew in the perturbative regime. For closer binaries, the orbital-plane asymmetry is no longer captured by the linear picture.
\end{abstract}

\begin{keywords}
hydrodynamics -- methods: numerical -- binaries: close -- stars: interiors -- stars: rotation
\end{keywords}

\section{Introduction}

Binary stars, consisting of two stars orbiting around their common center of mass, provide a natural laboratory for the dynamics of rotating self-gravitating fluids.
Interactions between the primary and secondary stars can produce contact binaries, cataclysmic variables, X-ray binaries, stripped-envelope supernova progenitors, Type Ia supernova progenitors, and compact-object mergers (\citealp{DeMarcoIzzard2017Binary}; \citealp{RoepkeDeMarco2023CE}; \citealp{Liu2023SNIaReview}). 
Despite their importance, some of the most important phases remain poorly understood. 
The theoretical difficulty is that a transition from a detached binary to either a mass-transferring system, or a common-envelope, or a merging system couples stellar structure, orbital dynamics, thermal physics, and intrinsically multidimensional hydrodynamics; it cannot be determined from ordinary single-star evolution alone.

Two key processes are mass transfer and internal flow induced in the stellar interiors. Mass transfer not only determines  the masses of individual stars but also dictates the exchange of their spins and the orbital angular momentum, while internal flow redistributes entropy, chemical composition, and angular momentum inside each component. Double white-dwarf mergers, for example, can leave a cold degenerate core surrounded by a hot differentially rotating corona and a centrifugally supported disk . The fate of the remnant depends on heating by dissipation of differential rotation, angular-momentum transport, cooling, and possible nuclear burning \citep{Pakmor2012StellarGadget, Schwab2021WDRemnants}.
Non-degenerate stellar mergers, on the other hand, raise related issues: the remnant is expected to be thermally out of equilibrium and rapidly rotating, and its subsequent appearance depends on internal mixing and angular-momentum loss \citep{Sills2005BlueStragglers, Schneider2019MassiveMergers, MathieuPols2025BlueStragglers}.

Observations are now beginning to constrain these processes from yet another side.
Gaia has transformed the census of wide, resolved, astrometric, and spectroscopic binaries, providing masses and orbital  parameters  for binaries of different types
 \citep{GaiaCollaboration2023NSS, Chevalier2023GaiaMasses}. 
TESS, by contrast, is especially powerful for close systems: short-cadence eclipsing-binary catalogues and full-frame-image searches constrain short-period eclipsing and ellipsoidal binaries, including contact systems and tidally distorted binaries \citep{Prsa2022TESS, Green2023TESS}. More direct information on the gas flow in the binary is available for young systems, in which  cavities, streams, circumstellar disks, and non-axisymmetric structures are spatially resolved \citep{Ragusa2021IRAS04158}.
 For older systems, the population properties of white dwarfs, magnetic fields, spin periods, surface abundances, and peculiar spectral classes provide  constraints
on the past common-envelope and merger processes\citealp{GentileFusillo2021WDCatalog};\citealp{Ferrario2015MWD}; \citealp{Schwab2021WDRemnants};
\citealp{Kilic2024DAQ}. Although these data do not yet reconstruct the internal flow field directly,  they  restrict the binary structures, mass-transfer states, and remnant properties ; theoretical models must explain them.

The equilibrium approach has a long history and provides the immediate background for the present work. Self-consistent-field methods were developed for rapidly rotating stars, rings, and multi-body fluids, and were later extended to relativistic cases  \citep{OstrikerMark1968, HachisuEriguchiSugimoto1982, HachisuEriguchi1982, EriguchiHachisu1983a, EriguchiHachisu1983b, Hachisu1986, Hachisu1986b, Komatsu1989a, Komatsu1989b, EriguchiFutamaseHachisu1990, EriguchiHachisuNomoto1994, Yasutake2015, UryuEtAl2019}. Non-axisymmetric equilibria with internal motions and quasi-equilibrium binaries with internal circulation were also explored by Uryu and Eriguchi \citep{UryuEriguchi1996, UryuEriguchi1998a, UryuEriguchi1998b, UryuEriguchi1999, EriguchiUryu1999}. Baroclinic stellar models based on different formulations and the study of gravity darkening
 show that thermal structure and internal circulation should be treated as a part of the stellar configuration, not merely as post-processing corrections \citep{UryuEriguchi1994, UryuEriguchi1995, EspinosaLara2007, EspinosaLara2011, Espinosa2012, EspinosaLara2013, Fujisawa2015, Yasutake2016, Zorec2017, Okawa2023, Ogata2023}. 
The hydrodynamical approach is equally important as it connects configurations
obtained in the equilibrium approach to binary evolution.
Grid-based  simulations either on a fixed mesh or on a moving mesh have
revealed many aspects of common-envelope and mass-transfer dynamics,
which are intrinsically of multi-scale and
multi-physics nature\citep{RoepkeDeMarco2023CE}.
White-dwarf merger simulations show that the outcome depends on  the post-merger thermodynamic structure and angular-momentum distribution, as well as on subsequent viscous evolution and nuclear burning \citep{Pakmor2012StellarGadget, Schwab2021WDRemnants,
Liu2023SNIaReview}.
Non-degenerate stellar merger remnants are just analogous, since their subsequent evolutions are dictated by mixing, winds, magnetic torques, and possible disk
interactions \citep{Sills2005BlueStragglers, MathieuPols2025BlueStragglers}.
Smoothed Particle Hydrodynamics (SPH) offers a complementary Lagrangian approach well suited to large-amplitude deformation, free surfaces, tidal stripping, and binary mergers \citep{Price2012SPH}, but spurious transients in the  stellar and orbital configuration may affect the subsequent mass-transfer , internal circulation, and compromise the inference of remnant spin.

This paper is concerned with the internal flow in the binary components. We are particularly interested in the effect of orbital motion. We take a hybrid approach: while we perform time-dependent SPH simulations, the goal is to obtain flows in (quasi-)equilibrium in the individual stars in binary. Our numerical models are rather restricted: we
 do not model mass transfer, merger dynamics, radiative transport, nuclear burning,  a self-consistent thermal balance.
We construct  equal-mass,  binary configurations in a rotating frame ; the treatment of spin is neglected and the entropy profile in space is prescribed instead, and examine the resulting quasi-steady baroclinic flows. The SPH code used here is based on FDPS \citep{fdps}, and self-gravity is evaluated with its monopole tree rather than with a grid-based Poisson solver.

From a fluid-dynamical perspective, the problem is one of baroclinicity-driven secondary
circulation in a  self-gravitating fluid subject to tides and orbital motion.
Once the density and pressure gradients are misaligned, a large-scale double-cell circulation is expected in
the stellar interior. The question addressed here is how orbital motion biases  baroclinicity-driven circulations and produces an orbital-plane asymmetry.
The resulting  flow is compared on the equatorial section systematically
with Dean vortices: the analogy holds for the two-cell topology and its skewness, although the driving physics is different. Classical Dean flow is a viscous wall-bounded flow in a curved pipe \citep{Dean1928, Berger1983, Germano1982, Kao1987, Germano1989, LiuMasliyah1993, Yamamoto1994}, whereas the  circulation of our concern is  induced by imposed baroclinicity in a nearly inviscid self-gravitating fluid, and its skewness is selected by orbital motion.

The organization of this paper is as follows. Section 2 describes the numerical setup and the  prescription of the baroclinicity. Section 3 presents the quasi-steady SPH solutions in the slow-orbital-rotation regime. Section 4 develops a minimal linear analysis for the orbital bias. Section 5 discusses the close-binary regime. The final section summarizes the results and  the limitations.
Throughout this paper, physical quantities are normalized by $M_\odot = R_\odot = G = 1$.

\section{Numerical Setup and Relaxation toward Quasi-Steady States}
\subsection{Initial conditions}

We consider an equal-mass binary. Each star is initially constructed from a one-dimensional Lane-Emden solution.
Although real binaries generally have unequal masses, our primary aim is to examine equilibrium structures and internal flow patterns in the simplest symmetric setting. This restriction allows us to isolate the effects of imposed baroclinicity before moving on to more complicated configurations.

\begin{figure}
\includegraphics[width=4.2cm]{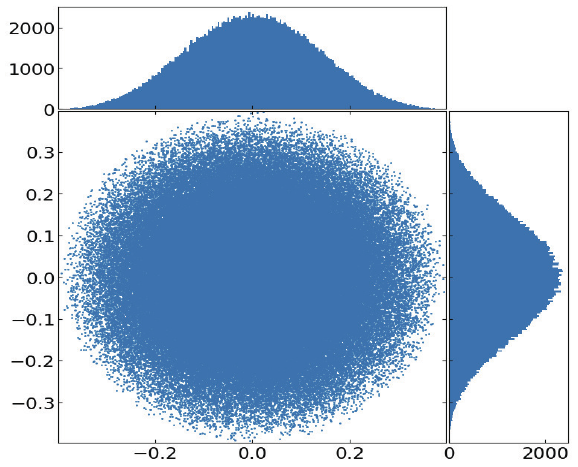}
\includegraphics[width=4.2cm]{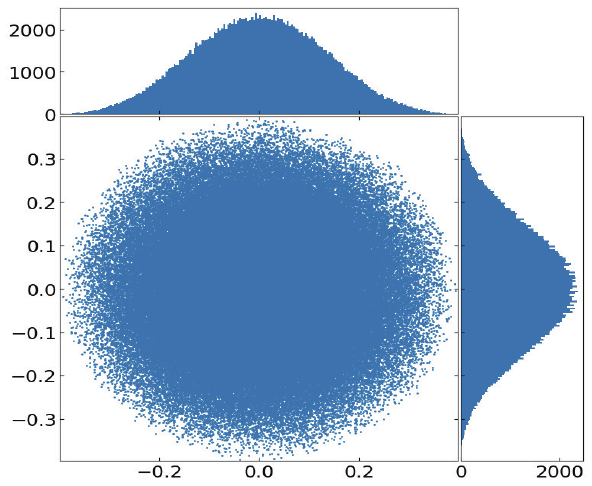}
\caption{The particle distribution in the $xy$-plane (left) and $xz$-plane (right) for one star with $10^5$ particles. The binary therefore contains $2 \times 10^5$ mass elements in this initial condition.}
 \label{fig:particle}
\end{figure}

The density profile is converted into an SPH particle distribution using von Neumann's rejection method \citep{Neumann1951}. In this procedure, particles are sampled randomly within a prescribed volume and are accepted or rejected according to the target density distribution. Regions of higher density therefore contain a larger number of accepted particles. The resulting particle distribution is shown in Fig.~\ref{fig:particle}, together with histograms on the $xy$- and $xz$-planes for one star with $10^5$ particles.

\blue{For the numerical simulations, we developed a density-based SPH code using the FDPS framework \citep{fdps}. The code employs a compact cubic-spline kernel and an adaptive smoothing length, $h=1.2(m/\rho)^{1/3}$. 
Artificial viscosity with the Balsara shear limiter \citep{Balsara1995} is applied 
only when particle pairs approach one another.
Self-gravity is calculated using the Barnes--Hut tree method \citep{BarnesHut1986} with the monopole approximation and Plummer softening ($\epsilon=0.01$). The equations of motion are integrated with a global time step determined by the Courant condition.}

Once the particle distribution is prepared, a trial orbital angular velocity and entropy are assigned to each particle. The stellar centers of mass orbit in the $xy$-plane. 

We impose circular orbits with a constant orbital angular velocity.

Even in this idealized setting, binary equilibria are not sufficiently understood, especially for baroclinic configurations. 
In realistic close systems, the density near the orbital axis may be lower owing to tidal deformation, and a temperature asymmetry associated with gravity darkening may then appear. In this paper, we do not attempt to compute that thermal structure self-consistently. 
Instead, we prescribe a spatial entropy profile to represent an imposed thermal asymmetry and examine the resulting quasi-steady flow in the co-rotating frame.

In this study, we employ a polytropic equation of state~(EOS):
\begin{equation}
p = \kappa \rho^\Gamma,
\end{equation}
Here, $p$, $\rho$, and $\Gamma$ denote the pressure, density, and polytropic exponent, respectively; in the present model, $\Gamma$ is also the adiabatic index. 

The coefficient \(\kappa\) specifies the prescribed spatial entropy profile.
We consider two cases: a barotropic case with constant $\kappa$, as commonly adopted in previous equilibrium studies, and a baroclinic case in which $\kappa$ varies spatially. Throughout this paper, we adopt the polytropic index $n=1.5$, corresponding to $\Gamma=1+1/n=5/3$.

For the barotropic case, we set $\kappa=\kappa_0$ with $\kappa_0=0.171$. For the baroclinic case, we assume
\begin{equation}
\kappa = \kappa_0 ( 1 + c\chi^2 ),
\label{eq:Kpro}
\end{equation}
where the coefficient $c$ is set to 0.25, and $\chi$ is defined as
\begin{equation}
\chi = \frac{\varpi-\varpi_{\rm min}}{\varpi_{\rm max}-\varpi_{\rm min}},
\end{equation}
where $\varpi$ is the distance from the orbital rotation axis, and $\varpi_{\rm max}$~($\varpi_{\rm min}$) is the maximum~(minimum) value of $\varpi$ within each star. 
This profile is used as a phenomenological proxy for a gravity-darkened temperature
distribution in which regions closer to the orbital axis are cooler\citep{Espinosa2012}.

\subsection{Relaxation Procedure and Diagnostics}
After assigning a trial angular-velocity profile and a $\kappa$~(entropy) distribution, we iteratively search for quasi-steady configurations through the following procedure:
\begin{itemize}
\item The momentum of every mass element in the $z$ direction is periodically removed. This extracts energy from the system while preserving the total angular momentum about the $z$ axis.
\item At regular intervals, we reset the angular velocity about the $z$ axis to a uniform value. This value is chosen so that the total angular momentum $L_z$ about the $z$ axis is conserved. In this reset step, we impose a non-spinning configuration for each component star, so that the remaining uniform rotation represents only the binary orbital motion. 
\item For the baroclinic models, the prescribed entropy profile is reset at every time step in order to maintain the imposed gravity-darkening-like thermal asymmetry.
\item Small numerical fluctuations in the total linear momentum are removed as needed.
\item The total energy is monitored throughout the relaxation process. After the configuration approaches a quasi-steady state, this quantity converges to an approximately constant value.
\item The virial quantity is also monitored. In exact equilibrium it should vanish, and here it is used as a diagnostic of
steadiness.
\item Once the total energy and virial quantity have converged, we continue the simulation without the first and second procedures and follow the corresponding dynamical quasi-steady state.
\end{itemize}

One possible concern is that the third procedure  be omitted at the last stage when seeking a true thermal-equilibrium solution.
 This is indeed a limitation of the present approach. Our baroclinic models are not fully self-consistent stellar equilibria, because the thermal structure is externally maintained rather than evolved with radiative transfer or nuclear burning. Accordingly, the solutions presented below should be interpreted as forced quasi-steady states under an imposed entropy distribution. We nevertheless adopt this procedure because it allows us to isolate the fluid response to a prescribed baroclinic forcing and to identify the resulting large-scale flow pattern.

The convergence of the total energy \(E_{\rm tot}\), defined below, provides the most direct measure of
steadiness.

For fixed conserved quantities, an equilibrium is a stationary point of the total energy, and a mechanically stable equilibrium is a local energy minimum. During relaxation, dissipation reduces \(E_{\rm tot}\); its convergence therefore indicates approach to a mechanically relaxed quasi-steady state.
\begin{eqnarray}
E_{\rm tot} = K+W+\frac{T}{\Gamma-1},
\end{eqnarray}
where $K$ and $W$ are the kinetic and gravitational energies, respectively, and $T$ is the pressure integral. These quantities are evaluated as sums over SPH particles:
\begin{eqnarray}
K &=& \frac{1}{2} \sum_{i}^N |{\bf v}_i|^{2} dm_{i},  \\
W &=& - \frac{1}{2} \sum_i^N \sum_{j\neq i}^N  \frac{dm_{i}~ dm_{j} }{| {\bf r}_i - {\bf r}_j  |},
\label{eq:W}\\
T &\equiv & \int p\,dV = \sum_{i}^N  \frac{p_i}{\rho_i}dm_{i}.
\end{eqnarray}
Here, ${\bf v}_i$, ${\bf r}_i$, $p_i$, and $\rho_i$ denote the velocity, position, pressure, and density of the $i$th particle. The factor of $1/2$ in Eq.~(\ref{eq:W}) avoids double counting of particle pairs. All particles are assigned the same mass, $dm_i=M/N$. For the $\Gamma$-law EOS adopted here, the internal energy is $U=T/(\Gamma-1)$; thus, $T$ is proportional to, but is not itself, the thermal energy. The spatial variation of $\kappa$ does not alter this instantaneous relation, although resetting $\kappa$ supplies or removes thermal energy and therefore breaks energy conservation in the baroclinic runs. By gradually removing energy while conserving the total angular momentum in the barotropic relaxation, we obtain a mechanically relaxed mass configuration. For the baroclinic models, the same mechanical relaxation is used, and the measured angular-momentum drift is quantified in Fig.~\ref{fig:amom}. 

During the barotropic relaxation, energy removal at fixed total angular momentum drives the system toward a local minimum of $E_{\rm tot}$. Once this mechanically relaxed state is reached, $E_{\rm tot}$ becomes nearly constant.

As an additional diagnostic of dynamical stability, we introduce the virial quantity
\begin{equation}
V_C =  \left |  \frac{2K+W+3T}{W} \right |.
\end{equation}
which vanishes in equilibrium by the virial theorem. This quantity is commonly used in the literature \citep[e.g.][]{Garcia2020}. The barotropic wide-binary and close-binary models reach $V_C=O(10^{-3})$, whereas the forced baroclinic wide-binary model remains at $O(10^{-2})$ or below because the imposed $\kappa$ profile is repeatedly reset.

We therefore use the convergence of $E_{\rm tot}$ and $V_C$ as the primary criteria for identifying quasi-steady states, and monitor the binary orbital radius $r_G$ and orbital angular velocity $\Omega$ as auxiliary diagnostics.
We define the representative orbital angular velocity as $\Omega=L_{z,1}/I_{zz,1}$, where $L_{z,1}$ and $I_{zz,1}$ are evaluated for Star-1 about the binary center of mass. For the present nearly non-spinning configurations, this quantity agrees with the center-of-mass estimate $\Omega_{\rm cm}=({\boldsymbol r}_G\times{\boldsymbol v}_G)_z/r_G^2$ to within about $0.2\%$ in the snapshots examined. We therefore use $\Omega$ as the orbital angular-velocity diagnostic throughout this paper. For the equal-mass models, the value inferred from either component is the same by symmetry, and, when the circular-orbit reset is applied, it coincides with the imposed uniform orbital angular velocity.
If the relaxation reaches a quasi-steady circular configuration, $r_G$ and $\Omega$ should also settle after $E_{\rm tot}$ and $V_C$ have converged.

\section{Slow-Orbital-Rotation Regime}
\subsection{Diagnostics of Quasi-Steady States}

First, we consider the barotropic cases, which have been relatively well researched in previous studies.
In particular, \citet{Garcia2020} conducted SPH calculations to determine equilibrium states of self-gravitating rotating bodies, including compact binaries, with barotropic EOSs.

\begin{figure}
\includegraphics[width=8.5cm]{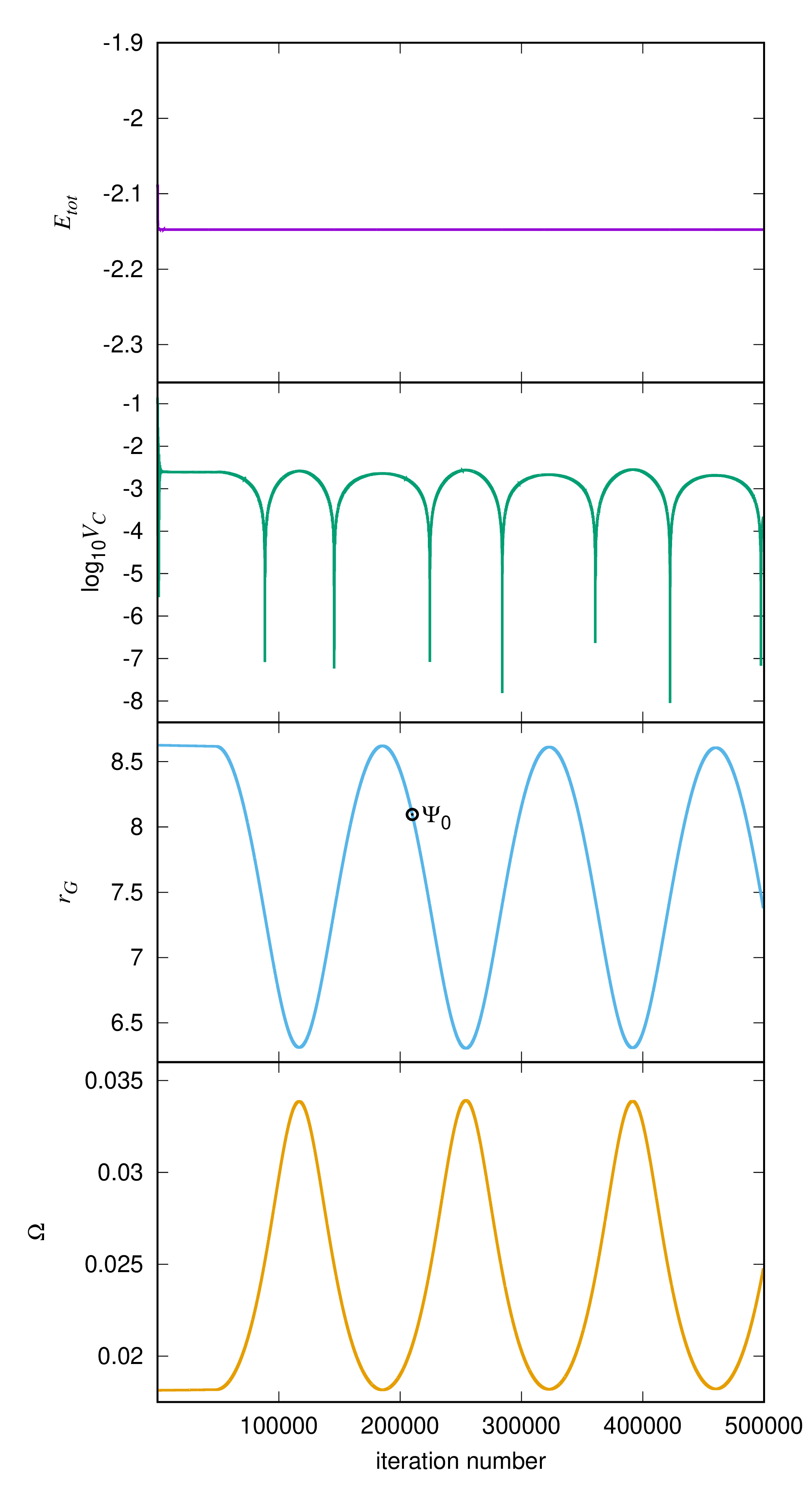}
\caption{\label{fig:4_brtr} Time evolution of four diagnostics for the barotropic model during relaxation and subsequent free evolution: the total energy $E_{\rm tot}$, the virial residual $V_C$, the orbital distance from the center of mass $r_G$, and the orbital angular velocity $\Omega$. Point $\Psi_0$ denotes the data used later for comparing flow states.}
\end{figure}

Following the setup described above, we construct two spherical stars using the Lane-Emden equation and search for relaxed binary configurations with the procedure outlined in the preceding section. As shown in Fig.~\ref{fig:4_brtr}, we focus on four diagnostics of the quasi-steady state, $E_{\rm tot}$, $V_C$, $r_G$, and $\Omega$. Until the structural changes settle, the velocity in the direction perpendicular to the orbital plane, i.e., the $z$-directional velocity, is reduced every 200 steps. This operation removes part of the mechanical energy. In the barotropic run, the subsequent angular-velocity reset is performed at fixed total angular momentum: all SPH particles are assigned a uniform orbital angular velocity about the $z$-axis, chosen from the current angular momentum, while the component stars are treated as non-spinning. The binary orbit is assumed to be circular only for this reset. This procedure is continued for 50,000 steps. Thereafter, we perform dynamic calculations without any further intervention and track the subsequent temporal evolution.
 
\begin{figure}
\includegraphics[width=8.5cm]{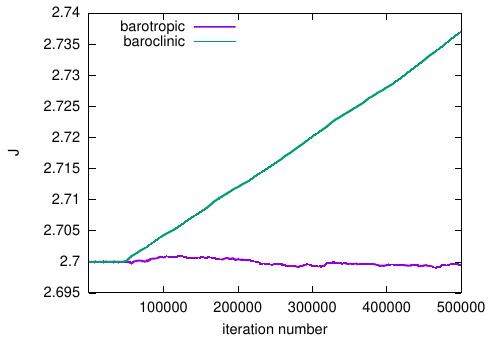}
\caption{\label{fig:amom}Time evolution of the total angular momentum in the barotropic and baroclinic wide-binary runs initialized at $J=2.7$. The barotropic model conserves $J$ to high accuracy during the relaxation and the subsequent free evolution, confirming that the energy-extraction procedure changes the orbital energy while leaving $J$ nearly unchanged. In the baroclinic model, by contrast, $J$ drifts by about $1.5\%$ by the final time, so the fixed-$J$ Keplerian estimate in the text is applied only to the barotropic run.}
\end{figure}

Excluding the initial thousands of steps during which the system relaxes toward a steady solution,
we find that by around 50,000 steps the four quantities have settled to approximately constant values. The angular momentum is conserved to a relative accuracy of $\sim10^{-5}$ in the barotropic run, as shown explicitly in Fig.~\ref{fig:amom}. After the periodic resets are stopped (after 50,000 steps), the binary system completes two orbital revolutions by 500,000 steps. This is reflected in the behavior of $r_G$ and $\Omega$.
As described in the previous section, up to 50,000 steps the orbital motion is periodically reset so that the binary follows a common circular orbit, and indeed the four quantities appear to have settled to constant values. However, once the system is allowed to evolve freely without further resets, the orbit deviates from a perfect circle and transitions into two elliptical orbits. 

In a bound two-body orbit, the eccentricity is  obtained by the orbital energy and angular momentum, since $E_{\rm orb}=-GM\mu/(2a)$ and $J_{\rm orb}^2=GM\mu^2 a(1-e^2)$. Equivalently,
\begin{equation}
e^2 = 1 + \frac{2E_{\rm orb}J_{\rm orb}^2}{G^2M^2\mu^3}
      = 1 - \frac{E_{\rm orb}}{E_{\rm circ}(J_{\rm orb})},
\end{equation}
where $E_{\rm circ}(J_{\rm orb})=-G^2M^2\mu^3/(2J_{\rm orb}^2)$ is the orbital energy of the  circular orbit with the same angular momentum. For the present equal-mass binary, $M=2m$ and $\mu=m/2$, where $m$ is the mass of each component star and $a$ is the semi-major axis. The total energy $E_{\rm tot}$ monitored in Fig.~\ref{fig:4_brtr} contains the self-gravitational and internal energies of both stars and is not substituted into this point-mass eccentricity relation. Instead, we estimate $E_{\rm orb}$ and $J_{\rm orb}$ independently from the center-of-mass variables $r_G$ and $\Omega$. In the barotropic relaxation procedure, energy is extracted while the total angular momentum is conserved in order to obtain a steady configuration. The final eccentricity is therefore not prescribed independently, but is inferred from the post-relaxation center-of-mass motion. For the Keplerian estimate below, we identify the displayed $\Omega$ with $\Omega_{\rm cm}$; their measured difference changes the estimated eccentricity by less than $0.004$ and is negligible compared with the quoted value $e\sim0.14$. These formulae are used only as point-mass Keplerian estimates for the center-of-mass motion. 
The SPH stars have finite size, self-gravity, internal energy, residual internal flow, numerical fluctuations associated with the finite particle representation, and tidal deformation, so the  point-mass estimate is intended only as a reference and does not
exactly represent the total energy of the binary.
Nevertheless, if we focus only on the center-of-mass motion of the two stars, the system may be approximated as two point masses with $m=1$ separated by $2r_G$. In the units adopted here, $G=m=1$, so $M=2$ and $\mu=1/2$. If the radial center-of-mass velocity is further neglected at the end of the relaxation stage, the relative orbital speed is $v_{\rm rel}\simeq 2r_G\Omega$, and the corresponding point-mass Keplerian estimates become
\begin{eqnarray}
E_{\rm orb}^{\rm K} &\simeq& r_G^2\Omega^2-\frac{1}{2r_G},\\
J_{\rm orb}^{\rm K} &\simeq& 2r_G^2\Omega .
\end{eqnarray}
This run is performed at a nominal fixed total angular momentum $J=2.7$ while energy is extracted during the relaxation, as confirmed by the barotropic curve in Fig.~\ref{fig:amom}. Around the end of the relaxation stage, Fig.~\ref{fig:4_brtr} gives $r_G\simeq 8.5$ and $\Omega\simeq 0.0187$. These center-of-mass values yield $J_{\rm orb}^{\rm K}\simeq 2.70$ and $E_{\rm orb}^{\rm K}\simeq -3.36\times 10^{-2}$. With $G=1$, $M=2$, and $\mu=1/2$, the eccentricity relation reduces to $e^2\simeq 1+4E_{\rm orb}^{\rm K}(J_{\rm orb}^{\rm K})^2$, giving an effective eccentricity $e\simeq 0.14$, consistent with the radial-variation estimate $e_r\sim0.15$ below. As an independent and more direct estimate from the simulation output, we also use the variation of the stellar center-of-mass distance,
\begin{equation}
e_r \simeq \frac{r_{G,\max}-r_{G,\min}}{r_{G,\max}+r_{G,\min}},
\label{eq:e_radial}
\end{equation}
where the factor of two between $r_G$ and the binary separation cancels for an equal-mass binary. From Fig.~\ref{fig:4_brtr}, $r_G$ varies roughly between $6.3$ and $8.6$, giving $e_r \sim 0.15$ for the freely evolving wide-binary orbit.

In the barotropic run, the total energy and angular momentum are conserved after the periodic resets are stopped. Nevertheless, residual particle-scale motion and finite-surface noise can still perturb the center-of-mass orbit. We therefore regard the small eccentricity as a residual of the constructed quasi-steady configuration, rather than as an independently prescribed orbital parameter.

Throughout this barotropic calculation, $V_C$ remains of order $10^{-3}$, satisfying the quasi-steady accuracy criterion adopted here. This value indicates that our configurations are less tightly virialized than those reported by \citet{Garcia2020}. This difference may partly reflect methodological and resolution differences, including their treatment of the stellar surface and the larger particle numbers used in several of their models. Such improvements will be a subject of future work. In this study, we focus on identifying quasi-steady states through dynamical calculations with this level of accuracy.

We mark this configuration as $\Psi_0$ in the figure, which will be used for comparison with the baroclinic case. The reason for choosing this particular time step for the comparison will be discussed later.

\begin{figure}
\includegraphics[width=8.5cm]{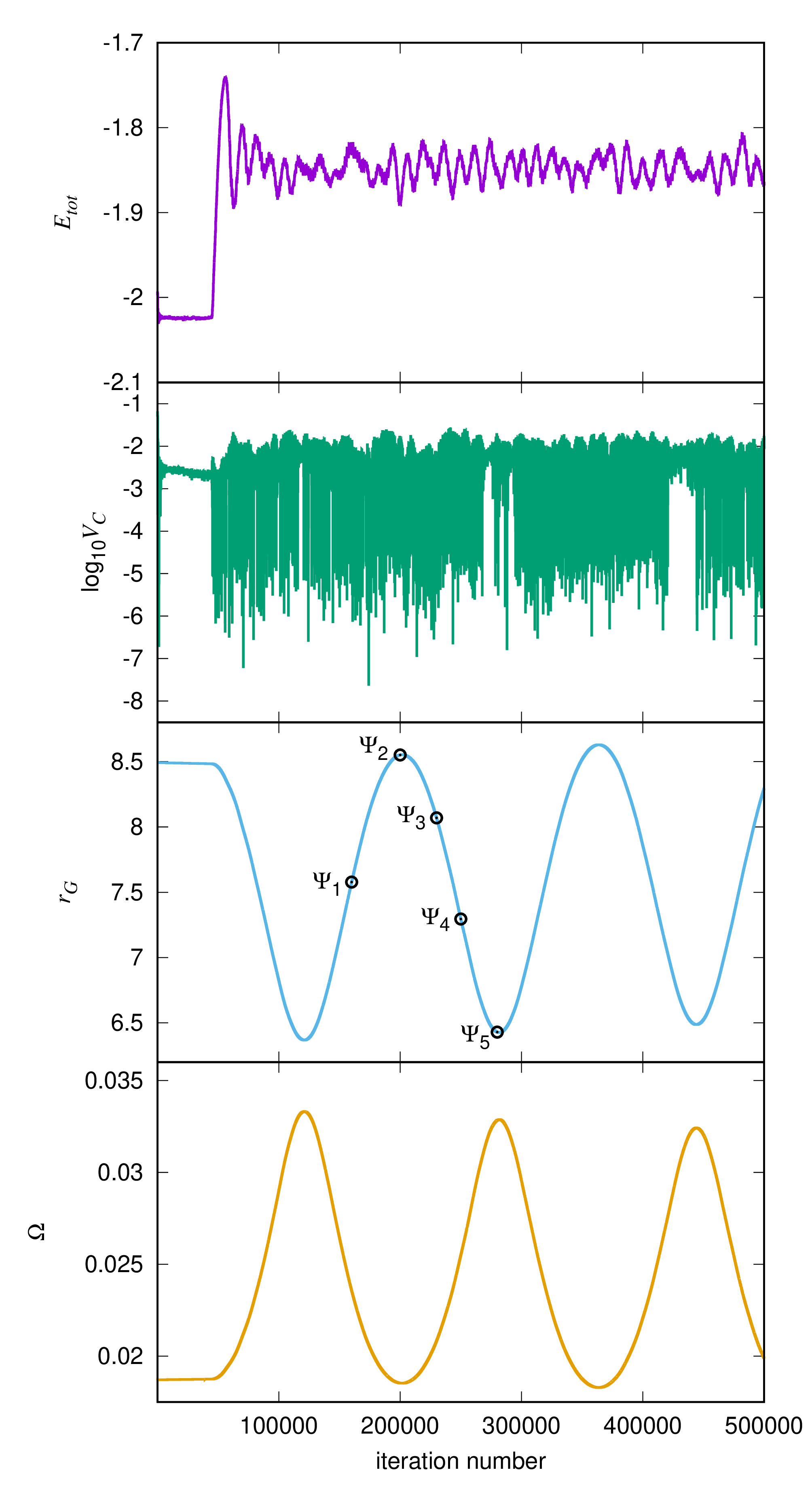}
\caption{\label{fig:4_brcl} Time evolution of the same four diagnostics as in Fig.~\ref{fig:4_brtr}, but for the baroclinic model during relaxation and subsequent free evolution. Points $\Psi_i$ ($i = 1, 2, 3, 4, 5$) indicate the data points used later to examine the configurations.}
\end{figure}

\bigskip

For the baroclinic case, we examine the same four diagnostics in Fig.~\ref{fig:4_brcl}.
The $z$-directional velocity is reduced every 200 steps until the structural changes settle, so that $E_{\rm tot}$ decreases. Unlike the barotropic case, however, the total angular momentum is not conserved to the same accuracy in the baroclinic run: Fig.~\ref{fig:amom} shows a drift of about $1.5\%$ by the final time. This procedure is again continued for 50,000 steps. In the baroclinic case, however, exact energy conservation is broken by the reset of the temperature distribution in Eq.~(\ref{eq:Kpro}) at each time step. Consequently, $E_{\rm tot}$ is offset between Figs.~\ref{fig:4_brtr} and~\ref{fig:4_brcl}. The same applies to $V_C$: its value remains of order $10^{-2}$ or below, while both $E_{\rm tot}$ and $V_C$ exhibit discontinuities because of the resetting procedure for the $\kappa$ profile.
From these results, the baroclinic runs should not be regarded as perfectly equilibrated solutions, but rather as forced quasi-steady states under the imposed $\kappa$ profile. Nevertheless, at late times the large-scale flow pattern becomes persistent, which is sufficient for the qualitative comparisons performed below. We note again that a more realistic treatment would require thermal evolution including radiative transport and nuclear burning, which is beyond the scope of the present study.

Regarding $r_G$ and $\Omega$, their behavior is similar in the two cases. The data points used later for examining the detailed structures are indicated in the figure as $\Psi_i$ ($i = 1, 2, 3, 4, 5$). Among these, the central data point $\Psi_3$ corresponds to approximately the same orbital phase as $\Psi_0$ in Fig.~\ref{fig:4_brtr}.

\subsection{Flow Patterns in the Rotating Frame}

\begin{figure}
\begin{center}
\includegraphics[width=5.3cm]{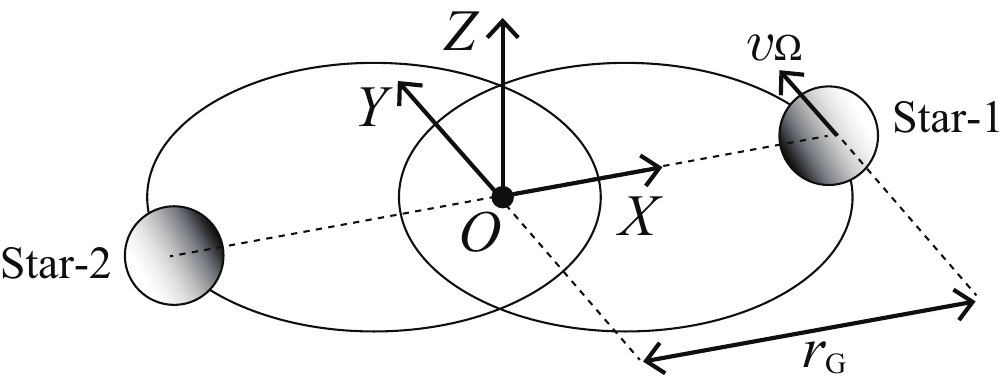}\\
\end{center}
\caption{Schematic illustration of the general orbital motion of an equal-mass binary system, with the origin placed at the center of mass of the system.
We adopt a co-rotating frame with the orbital angular velocity. In this frame, the $X$-axis is defined as the direction from the origin toward the center of mass of Star-1, and the $Y$-axis is defined along the direction of the orbital velocity, such that the orbit lies in the $XY$-plane. }
\label{fig:binary_orbit}
\end{figure}

In this subsection, we examine the motion of mass elements in the orbital co-rotating frame, in which the dominant orbital motion has been subtracted. In this frame, the centers of mass of the two stars lie along the $X$-axis, while the $Y$-axis is defined along the direction of orbital motion. A small residual radial motion remains, but it is negligible compared with the internal flow pattern discussed below. The geometry is summarized in Fig.~\ref{fig:binary_orbit}.

\begin{figure}
\includegraphics[width=7.9cm]{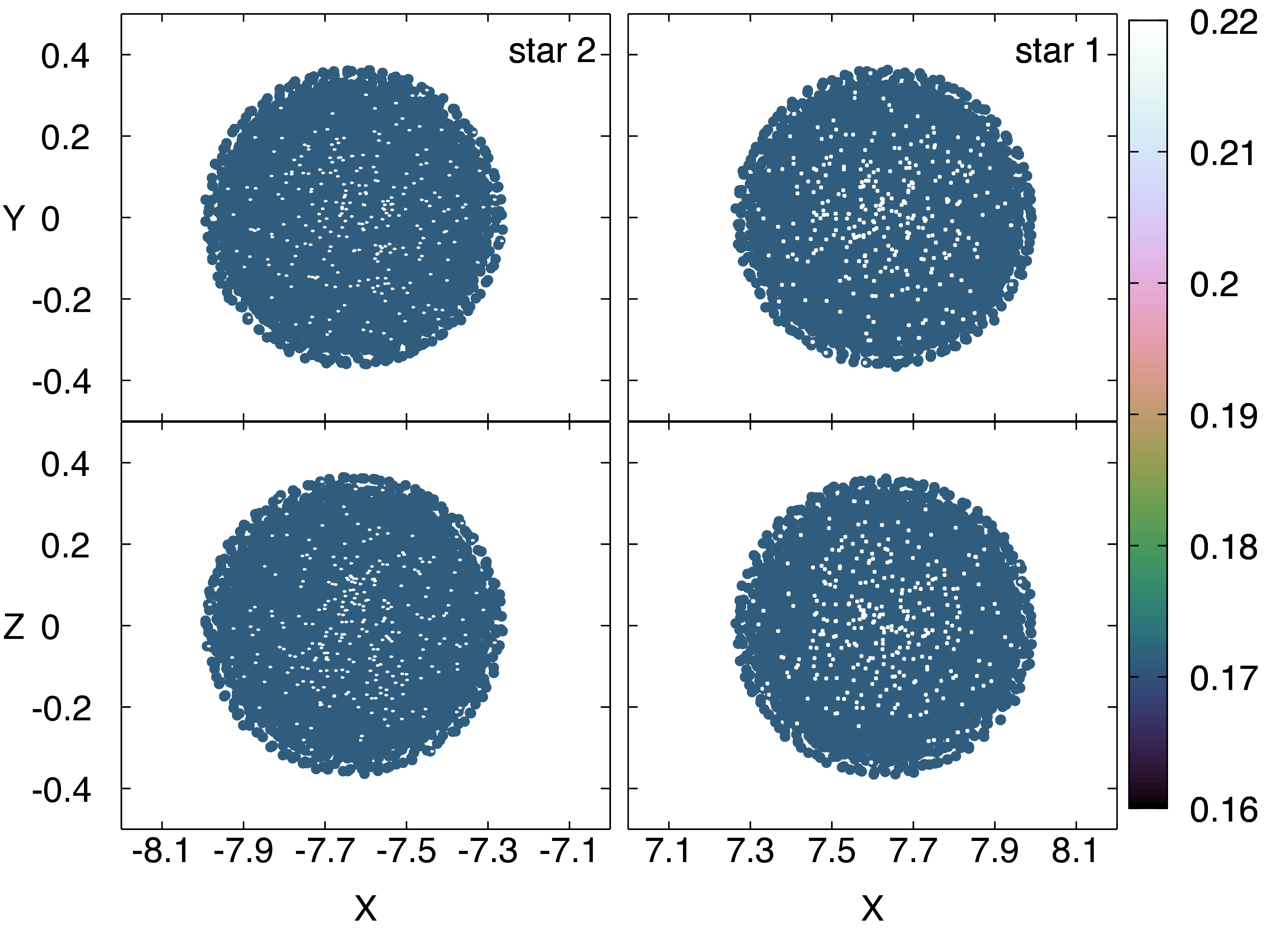}
\caption{\label{fig:brtr_K} Resulting $\kappa$ and velocity profiles for the barotropic case at $\Psi_0$ in Fig.~\ref{fig:4_brtr}. The $\kappa$ profile is shown by colors, while the velocity field is shown by arrows. The same arrow scale is used as in Fig.~\ref{fig:brcl_K}.}
\end{figure}

\begin{figure}
\includegraphics[width=7.9cm]{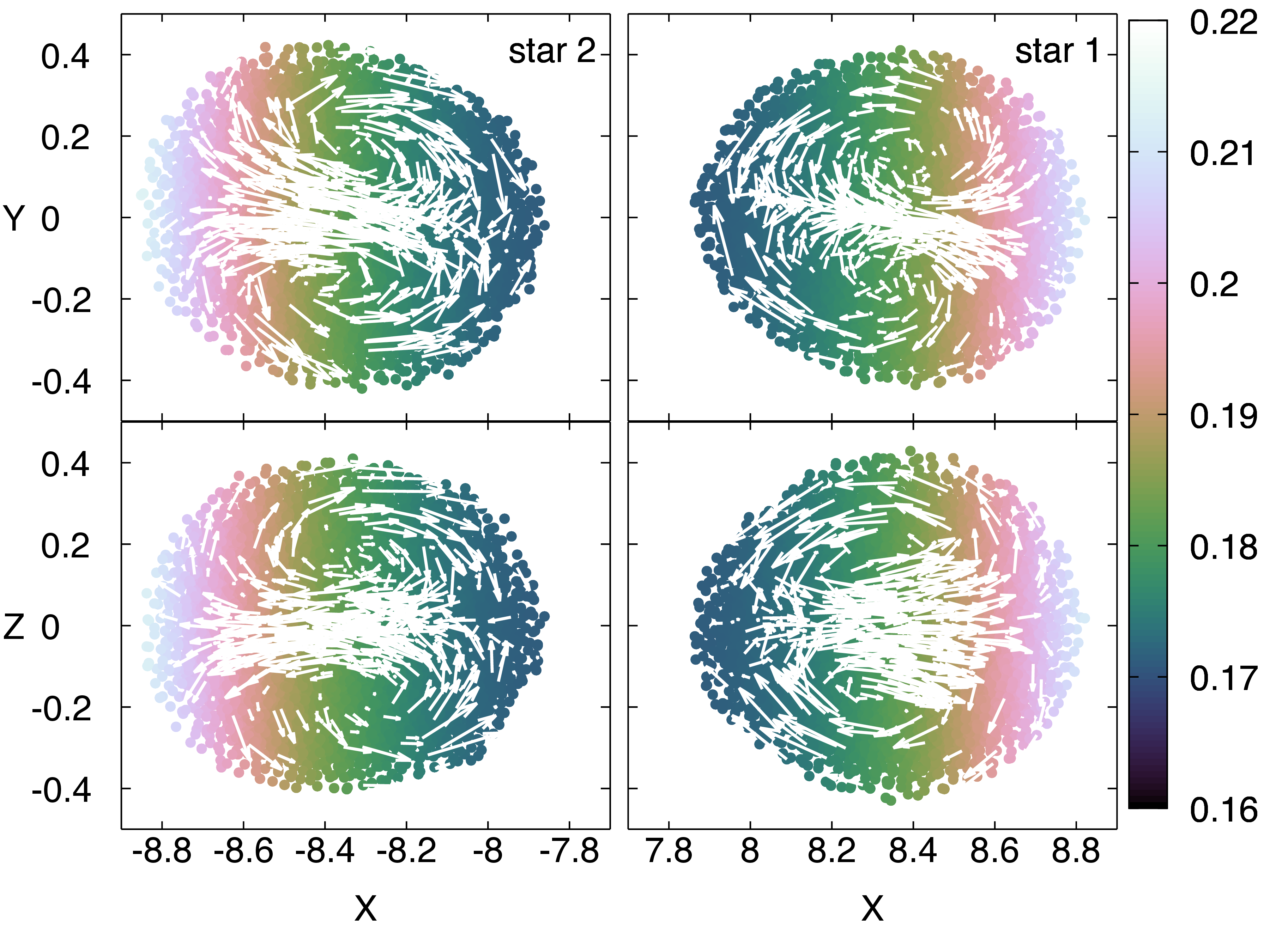}
\caption{\label{fig:brcl_K} Resulting $\kappa$ and velocity profiles for the baroclinic case at $\Psi_3$ in Fig.~\ref{fig:4_brcl}. The format and arrow scale are the same as in Fig.~\ref{fig:brtr_K}. }
\end{figure}

In Figs.~\ref{fig:brtr_K} and~\ref{fig:brcl_K}, we show slices of the $\kappa$ distribution on the $XY$- and $XZ$-planes for the barotropic case at phase $\Psi_0$ in Fig.~\ref{fig:4_brtr} and for the baroclinic case at phase $\Psi_3$ in Fig.~\ref{fig:4_brcl}. These phases are chosen to correspond approximately to the same orbital phase. The color contours show $\kappa$, while the white arrows represent the velocity field. The same arrow scale is used in both figures so that the flow amplitudes can be compared directly.

\begin{figure*}
\includegraphics[width=\textwidth]{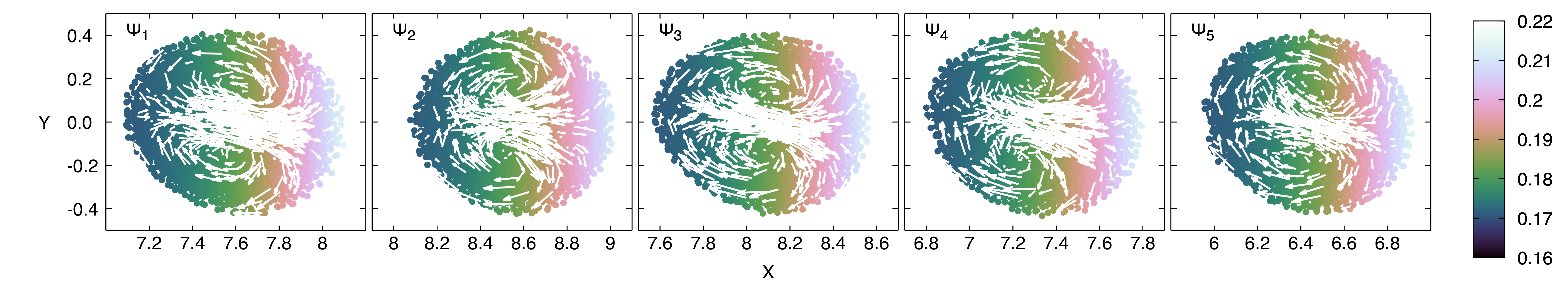}
\caption{\label{fig:phase} Resulting $\kappa$ and velocity profiles on the $XY$-plane of Star-1 for the baroclinic case.
$\Psi_i$ ($i = 1, 2, 3, 4, 5$) correspond to the orbital phases indicated in Fig.~\ref{fig:4_brcl}. }
\end{figure*}

We first summarize the numerical facts visible in Figs.~\ref{fig:brtr_K},~\ref{fig:brcl_K}, and~\ref{fig:phase}, and then turn to a physical interpretation based on the vorticity equation.


The numerical results are as follows. 
In the barotropic case, the velocity field in the co-rotating frame is weak and no coherent large-scale double-vortex pattern is identified. In contrast, the baroclinic case develops two global vortices on both the $XY$- and $XZ$-planes. On the $XZ$-plane, the two vortices are nearly reflection-symmetric about $Z=0$. On the $XY$-plane, in contrast, the circulation is not reflection-symmetric about the binary axis $Y=0$: the two cells are skewed in the same $Y$ direction. Fig.~\ref{fig:phase} shows that this skewness persists through the sampled orbital phases, although the detailed flow field changes slightly.

The equation of motion in the rotating frame is written as
\begin{eqnarray}
\frac{D{\boldsymbol v}}{Dt} + 2{\boldsymbol \Omega} \times {\boldsymbol v}= -\frac{{\boldsymbol  \nabla} p}{\rho}-{{\boldsymbol  \nabla} \Phi_{\rm eff}}
\label{eq:eom}
\end{eqnarray}
Here, ${\boldsymbol v}$, ${\boldsymbol \Omega}$, and $\Phi$ denote the velocity in the rotating frame, the angular velocity of the frame, and the gravitational potential, respectively. The effective potential is defined by
\begin{eqnarray}
\Phi_{\rm eff} \equiv \Phi-\frac{1}{2}({\boldsymbol \Omega}\times{\boldsymbol r})\cdot({\boldsymbol \Omega}\times{\boldsymbol r}),
\end{eqnarray}
so that the centrifugal acceleration is represented by the gradient of a potential without absorbing it into the pressure. This expression remains valid when the density varies spatially. We neglect physical viscous diffusion, treating the stellar-interior flow as effectively inviscid on the scales considered here.

Applying this equation in the binary co-rotating frame, we focus on one component star, which we call Star-1, while the companion is denoted Star-2. In this frame, ${\boldsymbol \Omega}$ is the orbital angular velocity, ${\boldsymbol \Omega}=(0,0,\Omega)$, and the gravitational potential may be decomposed as $\Phi=\Phi_1+\Phi_2$.
To identify the local direction in which baroclinicity generates vorticity, we isolate the baroclinic contribution to the vorticity evolution:
\begin{eqnarray}
\frac{D{\boldsymbol \zeta}}{Dt}
&=&
\frac{{\boldsymbol  \nabla}\rho \times {\boldsymbol  \nabla}p }{\rho^2},
\label{eq:vortex}
\end{eqnarray}
Here, ${\boldsymbol \zeta}$ is the vorticity vector. Equation~(\ref{eq:vortex}) is used only to determine the local direction of baroclinic vorticity generation, not as a complete transport equation or a steady-state vorticity balance. In the barotropic case, the density and pressure gradients are aligned, whereas the imposed temperature contrast in the baroclinic case makes ${\boldsymbol \nabla}\rho \times {\boldsymbol \nabla}p$ nonzero. Its direction, together with mass conservation and the absence of flow through the stellar surface, is consistent with fluid elements moving along the outer layers toward the orbital axis, descending into the stellar interior, and returning away from the orbital axis, thereby producing the observed double-vortex circulation. This interpretation is qualitatively consistent with the baroclinic circulation discussed by \citet{Espinosa2012}. The orbital-plane asymmetry of the developed circulation is considered separately in Section~4.

In Fig.~\ref{fig:phase}, we again present the $\kappa$ and velocity profiles for the baroclinic case on the $XY$-plane of Star-1, shown at different phases of the orbital motion. Although the snapshots are not identical, they exhibit the same characteristic behavior: two global vortices that are slightly asymmetric with respect to $Y = 0$. In the next subsection, we use linear analysis to examine in more detail the relationship between the vortex asymmetry and the orbital motion. 

\section{Minimal Linear Model for Orbital Bias}

\subsection{Primary Flow: Symmetric Baseline}

The aim of this section is deliberately limited. We do not attempt to construct an analytic counterpart of the full SPH solution. Instead, we address a more restricted question: if a symmetric baroclinic double-cell circulation has already been established, in which direction does weak orbital rotation displace it on the orbital plane? The model below is therefore used only to identify the sign and geometrical sense of the orbital bias in the perturbative regime. To this end, we consider the equation of motion for an inviscid fluid in the rest frame of Star-1, with the origin placed at the center of mass of Star-1.
\begin{eqnarray}
\frac{D{\boldsymbol v}}{Dt} + 2{\boldsymbol \Omega} \times {\boldsymbol v}= -\frac{{\boldsymbol  \nabla} p}{\rho}-{{\boldsymbol  \nabla} \Phi_1} 
-{\boldsymbol \Omega} \times ({\boldsymbol \Omega} \times {\boldsymbol r})
\label{eq:eom_star1}
\end{eqnarray}
Here, ${\boldsymbol r}$ denotes the position vector measured from the center of mass of Star-1, ${\boldsymbol r}=(x,y,z)$. In other words, we shift the origin from the binary center of mass to the center of mass of Star-1 by ${\boldsymbol r}_1 = (r_G, 0, 0)$ in Eq.~(\ref{eq:eom}). This equation is obtained from the orbital co-rotating frame centered on the binary center of mass under the approximation that the gravitational force from Star-2 balances the centrifugal force associated with the orbital motion. We treat $r_G$ and $\Omega$ as fixed at representative values for the selected orbital phase, so the analysis should be understood as a local, instantaneous model that neglects their residual time dependence. The details of the derivation are given in the Appendix.

To isolate this effect, we make four simplifying assumptions. First, the orbital angular velocity is small, and only terms linear in $\Omega$ are retained. Second, the flow is treated as incompressible. 

Third, the unperturbed circulation is assumed to be axisymmetric about the $x$-axis, which connects the centers of mass of the two stars.
Fourth, the stream-function construction below is used as a convenient representation of the primary double-cell flow, rather than as a self-consistent solution of the baroclinic stellar structure. These assumptions  may be  restrictive, but they allow us to track how the Coriolis term biases an otherwise symmetric circulation.

At this point, one caveat should be stated explicitly. The imposed entropy distribution is motivated by gravity darkening, which is physically expected to be strongest in close binaries rather than in the widely separated systems considered in the perturbative regime. The purpose of the present linear analysis is not to model gravity darkening quantitatively in that regime, but to isolate the weak-orbital-bias response of an already established baroclinicity-driven  double-vortex flow. In this section, the  prescribed entropy
profile should therefore be understood as a phenomenological proxy for thermal asymmetry, introduced so that the effect of orbital motion can be extracted in the cleanest possible limit.

Neglecting terms of second order in $\Omega$, we transform the equation into spherical coordinates. The details of this derivation are also given in the Appendix. The angular convention used here is slightly different from the usual spherical polar coordinates. The polar axis is chosen to be the $x$-axis, so that $\theta=0$ corresponds to the positive $x$ direction. The angle $\phi$ is then the azimuthal angle around the $x$-axis, measured counterclockwise on the $yz$-plane when viewed from the positive $x$ direction. This coordinate choice is illustrated in Fig.~\ref{fig:coordinate}. The resulting components of the Euler equation are
\begin{align}
 {\mathcal D}v  - \frac{w^2+u^2 }{r}  - 2\Omega  ( w  \cos\phi  & -  u \cos\theta \sin\phi ) \nonumber \\
&= -\frac{1}{\rho}\frac{\partial p}{\partial r} - \frac{\partial \Phi_1}{\partial r} ,
\label{eq:NS_1} 
\\
{\mathcal D}w + \frac{vw- u^2 \cot{\theta}}{r}   + 2\Omega  ( v & \cos\phi  - u \sin\theta \sin\phi ) \nonumber  \\
&= -\frac{1}{\rho r}\frac{\partial p}{\partial \theta} -  \frac{1}{r}\frac{\partial \Phi_1}{\partial \theta} ,
\\
{\mathcal D}u  +\frac{vu + wu \cot{\theta} }{r}   +  2\Omega ( - & v \cos\theta \sin\phi  + w \sin\theta \sin\phi ) \nonumber \\
= -\frac{1}{\rho r \sin{\theta}} & \frac{\partial p}{\partial \phi} -  \frac{1}{r \sin{\theta} }\frac{\partial \Phi_1}{\partial \phi}, 
\label{eq:NS_3}
\end{align}
where ${\boldsymbol v}$ is the velocity in this coordinate system, ${\boldsymbol v}=(v_r,v_\theta,v_\phi)=(v,w,u)$, and ${\mathcal D}$ is the Lagrangian derivative defined as
\begin{eqnarray}
{\mathcal D} =   v\frac{\partial}{\partial r} + \frac{w}{r} \frac{\partial}{\partial \theta} + \frac{u}{r \sin{\theta}} \frac{\partial}{ \partial \phi} .
\label{eq:D}
\end{eqnarray}
Since we focus on quasi-steady states, we neglect the explicit time-derivative term.

For simplicity, we assume incompressible hydrodynamics with constant density $\rho_0$. We then normalize the pressure and the gravitational potential by the density and define the effective pressure as $p_e(r, \theta)  \equiv p /\rho_0 + \Phi_1$. The continuity equation in this coordinate system reads
\begin{eqnarray}
\frac{\partial v}{\partial r} + \frac{1}{r} \frac{\partial w}{\partial \theta} + \frac{1}{r \sin\theta} \frac{\partial u}{\partial \phi} + \frac{2 v}{r}+ \frac{w}{r} \cot{\theta} = 0
\label{eq:continuity}
\end{eqnarray}

The displacement of the two vortices is weak at large $r_G$, i.e. in the small-$\Omega$ regime, and becomes more readily measurable as $\Omega$ increases. We therefore allow $v$ and $w$ to depend on $\Omega$. We adopt a linear expansion of $v$ and $w$ in $\Omega$:
\begin{eqnarray}
v(r, \theta, \Omega) &=& \bar{v}(r, \theta)  +  \Omega \delta v(r, \theta)\\
w(r, \theta, \Omega) &=& \bar{w}(r, \theta) + \Omega \delta w(r, \theta),
\end{eqnarray}
Here, $\bar{v}$ and $\bar{w}$ denote the $\Omega$-independent primary flow, while $\delta v$ and $\delta w$ represent the $\Omega$-dependent corrections (secondary flow) that give rise to the vortex asymmetry.

For the primary (baseline) flow, we retain the lowest-order (zeroth-order) terms in $\Omega$ of Eq.~(\ref{eq:eom_star1}) and denote the baseline effective pressure simply by $p_e$. This yields
\begin{eqnarray}
\bar{v} \bar{v}_{,r} + \frac{ \bar{w} \bar{v}_{,\theta} -  \bar{w}^2 }{r}   &=& -p_{e,r}  \\
\bar{v} \bar{w}_{,r} + \frac{ \bar{w} \bar{w}_{,\theta}+   \bar{v} \bar{w} }{r} &=&  -\frac{p_{e,\theta}}{r}  \\
\bar{v}_{, r} + \frac{\bar{w}_{, \theta}+2\bar{v} + \bar{w} \cot{\theta}  }{r} &=& 0,
\label{eq:eom_primitive}
\end{eqnarray}
where commas denote partial derivatives with respect to the indicated variables (e.g. ``${,r}$" and ``${,\theta}$"). We also set the azimuthal velocity to zero, consistent with the numerical results.

We now introduce a stream function following \citet{Pekeris1980} and solve Eq.~(\ref{eq:eom_primitive}) in the form
 \begin{eqnarray}
  \PP{\psi}{r}  +  \frac{\sin \theta}{r^2} \P{}{\theta} 
 \left(\frac{1}{\sin \theta} \P{\psi}{\theta} \right) 
  = -\rho_0 r q(\psi) \sin \theta  \ ,
  \label{eq:stream}
 \end{eqnarray}
where $q$ is an arbitrary function of $\psi$. The stream-function formulation is introduced only as a compact way to represent the symmetric primary circulation.
In a baroclinic fluid, it is not generally guaranteed that the flow can be represented by Eq.~(\ref{eq:stream}) with $q$ as a function of $\psi$ alone. In this work, however, we assume that the departure from the barotropic case is sufficiently small for Eq.~(\ref{eq:stream}) to represent the primary flow.
We therefore use the following construction as a kinematic model for the SPH flow pattern, not as a complete equilibrium theory. Under the axisymmetric assumption, we represent the stream function by an expansion in associated Legendre functions, following \citet{Pekeris1980}:
\begin{eqnarray}
 \psi(r,\theta) &=& r \sin \theta \sum_{n=1}^{\infty}
 \frac{P_{2n-1}^1(\cos\theta)}{2n(2n-1)} \nonumber \\
 &&\times \int_0^{\pi/2} d\theta' \sin\theta'\,
 P_{2n-1}^1(\cos\theta') \nonumber \\
 &&\times \int_0^{r_s(\theta')} d r'\, r'^2
 f_{2n-1}(r, r') \rho_0 q(r',\theta') \nonumber  \\
 &+& r\sin \theta \sum_{n = 1}^{\infty}
 \alpha_{2n-1}P_{2n-1}^1(\cos \theta) r^{2n-1} ,~~
 \label{eq:Phi}
\end{eqnarray}
 where $P_\ell^1$ is the associated Legendre function of degree
$\ell$ and order $1$ and $r_s(\theta)$ expresses the shape of the deformed surface of the star.
The kernel function $f_n$ is given as
\begin{eqnarray}
   f_{n}(r,r')
 = \left\{
   \begin{array}{lr}
     r'^{n}/r^{n+1}  & (r \geq r') \\
     r^{n}/r'^{n+1}  & (r \leq r'),
   \end{array}
 \right.  
\end{eqnarray}
and $\alpha_n$ are coefficients. Since we have assumed that there is no external wind, the $r$ component of the velocity ($v_r$) must vanish at the surface.  Therefore, the boundary condition for the stream function on the surface is as follows:
\begin{eqnarray}
 \psi(r_s, \theta) = 0.
 \label{eq:boundary}
\end{eqnarray}
We fix the coefficients to fulfill this boundary condition \citep{EriguchiMueller1991, Fujisawa2015}.

The velocity components are expressed in terms of the stream function under axisymmetry as
\begin{eqnarray}
\bar v &=& \frac{1}{\rho_0 r^2 \sin \theta } \P{\psi}{\theta} ,\\
\bar w &=& -\frac{1}{\rho_0 r \sin \theta } \P{\psi}{r} .
\end{eqnarray}
Assuming that the entropy gradient is small, we may expand the stream function as follows

\begin{eqnarray}
  \psi(r,\theta) & = &   \psi^{(0)}(r,\theta) + \varepsilon \psi^{(1)}(r,\theta) ,
\end{eqnarray}
where $\varepsilon$ is a small parameter and
 $\psi^{(0)}(r,\theta)$ denotes the stream function without an entropy gradient and therefore vanishes in practice.
This implies that $\bar v$ and $\bar w$ arise from the stream function at first order in $\varepsilon$.
Accordingly, the  function $q$ is also expected to be linear in  $\varepsilon$, and we thus assume the following in Eq. (\ref{eq:stream}) form
\begin{eqnarray}
 q & = &   - \varepsilon \rho_0 r \sin{\theta}. 
\end{eqnarray}

We further assume that  $p_e$ is  expanded  to second order in $\varepsilon$
\begin{eqnarray}
  p_e(r,\theta) & = &   p_e^{(0)}(r,\theta) + \varepsilon^2 p_e^{(2)}(r,\theta).
\end{eqnarray}
Again, $p_e^{(0)}(r,\theta)$ represents the sum of the pressure term and the gravitational potential, $p^{(0)} /\rho_0 + \Phi^{(0)}$, without an entropy gradient.
 Assuming its spherical symmetry, we may write
\begin{eqnarray}
\frac{\partial p_e^{(0)}}{ \partial r}  &=& \frac{1}{\rho_0}\frac{\partial p^{(0)}}{\partial r}   + \frac{\partial \Phi^{(0)}}{\partial r}  = 0.
\end{eqnarray} 

 The stellar surface as a sphere  $r_s(\theta)=r_0$.
 Imposing the boundary condition (\ref{eq:boundary})  to first order in $\varepsilon$ , we set
$\alpha_{2n-1}^{(0)} = 0$, $\alpha_{1}^{(1)} = \frac{\alpha}{3} \rho_0^2$, and $\alpha_{2n+1}^{(1)} = 0$, where $\alpha$ is an arbitrary constant.  Then we obtain 
\begin{eqnarray}
 \psi^{(1)} &=& \frac \alpha 2  \rho_0^2  \sin^2 \theta r^2 (r^2 - r_0^2) ,\\
  \bar v  &=& - \alpha \rho_0 (r^2-r_0^2) \cos \theta, \label{vbar} \\
  \bar w &=&   \alpha \rho_0  (2 r^2-r_0^2) \sin \theta, \label{wbar}
\end{eqnarray}
where the signs of $\bar v$ and $\bar w$ are arbitrary but are chosen to be consistent with the numerical results.

These expressions are intended to represent, at a qualitative level, the two symmetric vortices obtained from the SPH fluid simulations.
For notational simplicity, we set $\bar v = g(r) \cos \theta$ and $\bar w = h(r) \sin \theta$.
Then, the Euler equation and the continuity equation become
\begin{eqnarray}
g g_{,r} \cos^2{\theta}  - \frac{ h g + h^2 }{r} \sin^2{\theta}  &=&   -p_{e,r} ^{(2)}  \label{eq:r_eq} ,\\
\left( g h_{,r} + \frac{ hg + h^2 }{r} \right) \sin{2\theta} &=&  - 2\frac{p_{e,\theta}^{(2)}  }{r}   \label{eq:theta_eq},\\
g_{,r}  + 2\frac{h + g }{r} &=& 0 .
\label{eq:continuous}
\end{eqnarray}
One can readily verify that the continuity equation (\ref{eq:continuous}) holds and that all quantities remain finite at $r=0$.
An expression for $p_e^{(2)}$  that automatically satisfies the  equations  (\ref{eq:r_eq}) and (\ref{eq:theta_eq}) simultaneously is given as 
\begin{eqnarray}
p_{e}^{(2)} = \frac{\alpha^2 \rho_0^2}{4} \left\{  r_0^2 r^2 - (2r^4-3r_0^2 r^2)\cos {2\theta}  \right\} .
\end{eqnarray}

\subsection{Secondary Flow: Vortex Displacement}

We aim to obtain a qualitative understanding of the physical mechanism responsible for the shift of the two vortices toward the colatitudinal direction in the orbital ($XY$-)plane. Hence, we set $\phi = 0$ and focus  on the $r$- and $\theta$-components of the Euler equation. To examine the secondary flow, we consider terms at first order in $\Omega$. The Euler equation at this order becomes
\begin{eqnarray}
\bar{v}\,\delta v_{,r} + \delta v\, \bar v_{,r} 
 + \frac{\bar w\, \delta v_{,\theta} + \delta w\, \bar v_{,\theta}
 - 2 \bar w \delta w}{r} - 2 \bar w = 0, \nonumber \\
\label{eq:eom_secondary1}
\\
\bar{v}\,\delta  w_{,r} + \delta v\, \bar w_{,r} 
 + \frac{\bar w\, \delta w_{,\theta}+\delta w\,\bar w_{,\theta}
 +\bar v \delta w + \bar w \delta v}{r} + 2 \bar v = 0 .
\nonumber \\
\label{eq:eom_secondary2}
\end{eqnarray}
 In the numerical results , the vortex core moves  predominantly in the $\theta$ direction; hence, the dominant contribution to  the angular velocity is inferred to be
\begin{eqnarray}
\delta v(r, \theta)  &=& G(r) \sin{ \theta },
\label{eq:secondary1}
 \\
\delta w(r, \theta) &=& H(r) \cos{ \theta}.
\label{eq:secondary2}
\end{eqnarray}
where $G(r)$ and $H(r)$ are  functions of $r$.

 Determining the radial functions $G(r)$ and $H(r)$ throughout the star would require solving Eqs.~(\ref{eq:eom_secondary1}) and~(\ref{eq:eom_secondary2}) with the other angular modes in addition to those retained in Eqs.~(\ref{eq:secondary1}) and~(\ref{eq:secondary2}), together with the continuity equation, the thermodynamic relation for the pressure perturbation, and appropriate boundary conditions. 
Here, however, we restrict the analysis to  specific regions within the stars and seek a qualitative explanation for the direction of the circulation's deviation from axisymmetry rather than a complete solution.

First, we focus on the center of the primary vortices and denote its coordinates by $(r_C, \theta_C)$, where $\bar v(r_C, \theta_C)=0$ and $\bar w(r_C, \theta_C)=0$.
At this position, Eqs.~(\ref{eq:eom_secondary1}) and~(\ref{eq:eom_secondary2}) give
\begin{eqnarray}
\delta v_C~ \bar v_{C,r} + \frac{  \delta w_C~ \bar v_{C,\theta}  }{r_C}  = 0,\\
\delta v_C~ \bar w_{C,r} + \frac{ \delta w_C~\bar w_{C,\theta} }{r_C}  = 0 .
\end{eqnarray}
where the subscript C denotes quantities evaluated at the primary-vortex center; thus $r_C = r_0 / \sqrt{2}$ and $\theta_C=\pi/2$.
We already know the analytic form of $\bar v$ and $\bar w$ from Eqs.~(\ref{vbar}) and~(\ref{wbar}), and hence we immediately obtain $\delta v_C = 0$ and $\delta w_C =0$. This result implies that the Coriolis force does not act directly at the vortex center, where the primary velocity vanishes, as expected for a stagnation point of the primary circulation. The same conclusion holds for the other vortex located in the $\phi = \pi$ plane.

\begin{figure}
\begin{center}
\includegraphics[width=8cm]{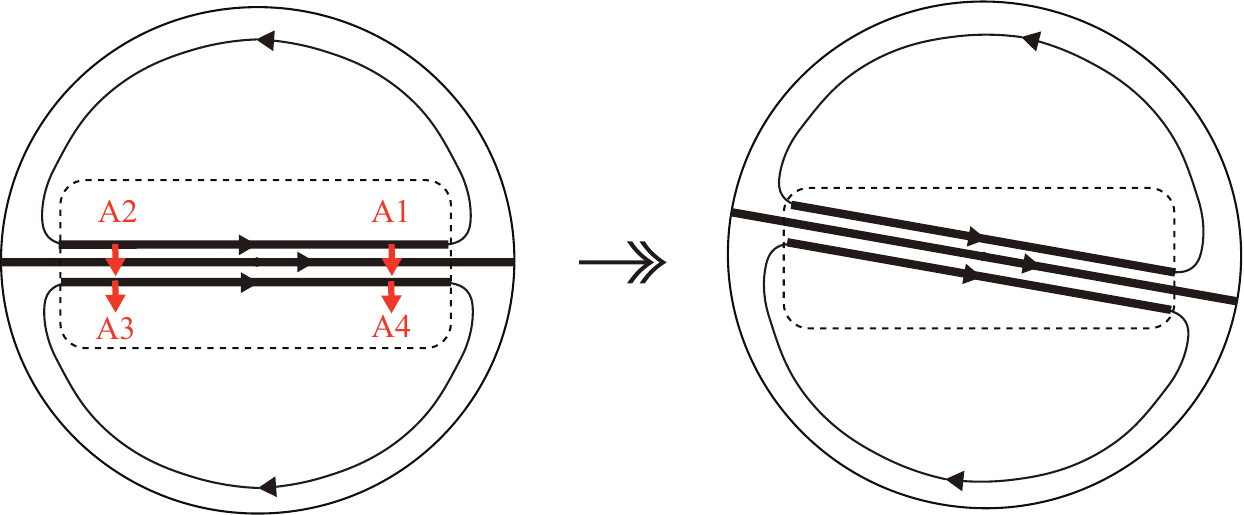}
\end{center}
\caption{\label{fig:center} Schematic illustration of the shift of the flow, $\delta w$, around $y = 0$ for Star-1, indicated by red vectors in the left panel.
The regions A1, A2, A3, and A4 are defined in the text and correspond to the first through fourth quadrants with respect to $y = 0$. In these regions, the radial component of the secondary flow vanishes, i.e. $\delta v=0$.
The right panel shows the resulting flow after the shift. Note that, because the $x$-axis is chosen as the polar axis, the positive $\theta$ direction in the cross section at $\phi=0$ ($y>0$) is opposite to that at $\phi=\pi$ ($y<0$).
}
\end{figure}

Let us  consider, instead , the region near $y = 0$, where the velocity is finite and the particle density is relatively high.  The influence of the Coriolis force is most pronounced. 
In particular, we focus on the second quadrant, which we call ``A2,''
corresponding to the region around $\theta_{A2}\sim\pi$ with
$\phi_{A2}=0$.  The primary velocity in the
$\theta$ direction is expected to be zero in this region:
$\bar w_{A2}=0$ (see the left panel in Fig.~\ref{fig:center}).
Then, Eq.~(\ref{eq:eom_secondary2}) yields
\begin{eqnarray}
\bar{v}_{A2} ~\delta  w_{A2,r} + \frac{ \delta w_{A2}~\bar w_{A2,\theta}  +\bar v_{A2} \delta w_{A2} }{r}  + 2 \bar v_{A2} = 0.~~
\label{eq:A2}
\end{eqnarray}
Since $\cos\theta_{A2}=-1$, Eq.~(\ref{eq:secondary2}) gives $\delta w_{A2}=-H_{A2}$. Substituting the explicit primary flow into Eq.~(\ref{eq:A2}) therefore gives
\begin{eqnarray}
H_{A2,r} - \frac{r_{A2}}{r_{A2}^2-r_0^2}H_{A2}-2=0.
\label{eq:A2_H}
\end{eqnarray}
The boundary condition follows from Eq.~(\ref{eq:A2}) itself. At the spherical surface $r_{A2}=r_0$, $\bar v_{A2}=0$ whereas $\bar w_{A2,\theta}=-\alpha\rho_0r_0^2\neq0$, and hence Eq.~(\ref{eq:A2}) requires $\delta w_{A2}(r_0)=0$. The corresponding condition on the amplitude is $\lim_{r_{A2}\to r_0^-}H_{A2}=0$ , in which , $r\to r_0^-$ denotes the limit toward the stellar surface from the interior, $r<r_0$. Because Eq.~(\ref{eq:A2_H}) is singular at $r_0$, this limiting value alone does not select a unique integral curve. We additionally require the radial derivative of the secondary velocity to remain finite as $r_{A2}\rightarrow r_0^{-}$. These surface regularity conditions select
\begin{eqnarray}
H_{A2}
=2\sqrt{r_0^2-r_{A2}^2}
\left[
\arcsin\left(\frac{r_{A2}}{r_0}\right)-\frac{\pi}{2}
\right].
\end{eqnarray}
Thus $H_{A2}<0$ for $0\leq r_{A2}<r_0$, and therefore $\delta w_{A2}=-H_{A2}>0$. 

The positive sign is consistent with the  Coriolis deflection toward the $\delta\theta>0$ direction.
 
Similarly, we now focus on the third quadrant and perform the same analysis.
Let us denote this area as ``A3,'' corresponding to the region around $\theta_{A3} \sim \pi$ with $\phi_{A3} = \pi$.
At $\phi_{A3}=\pi$, the Coriolis term changes sign. Repeating the same procedure gives
\begin{eqnarray}
H_{A3,r} - \frac{r_{A3}}{r_{A3}^2-r_0^2}H_{A3}+2=0.
\label{eq:A3_H}
\end{eqnarray}
The same surface conditions, $\lim_{r_{A3}\to r_0^-}H_{A3}=0$ and a finite radial derivative, select
\begin{eqnarray}
H_{A3}
=2\sqrt{r_0^2-r_{A3}^2}
\left[
\frac{\pi}{2}-\arcsin\left(\frac{r_{A3}}{r_0}\right)
\right].
\end{eqnarray}
This gives $H_{A3}>0$ and hence $\delta w_{A3}=-H_{A3}<0$. Although the positive $\theta$ direction is opposite in the cross sections at $\phi=0$ and $\phi=\pi$, this result represents the same  direction, again consistent  with the  Coriolis deflection as in A2.

Applying the same local calculation to A1 and A4 shows that the perturbations $\delta w_{A1}<0$ and $\delta w_{A4}>0$, respectively, arise in directions consistent with the Coriolis force.
As shown above, the Coriolis force corresponds to the terms
$-2\bar w$ and $+2\bar v$ in Eqs.~(\ref{eq:eom_secondary1}) and~(\ref{eq:eom_secondary2}), which are first order
in $\Omega$. These terms appear directly as the final terms, $-2$ and
$+2$, in Eqs.~(\ref{eq:A2_H}) and~(\ref{eq:A3_H}), respectively, and determine the sign of
the perturbation $\delta w$.
Taking account of the orientation of the spherical basis in each cross section, the four local deflections have the same physical direction near $y=0$. 
Fig.~\ref{fig:center} summarizes this common displacement and the resulting shift of the flow pattern.
They may be  considered to bias the double-cell pattern.
 
\begin{figure}
\includegraphics[width=4.2cm]{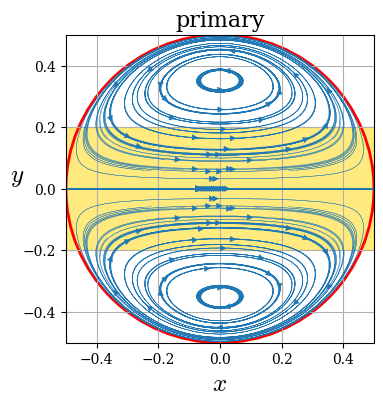}
\includegraphics[width=4.2cm]{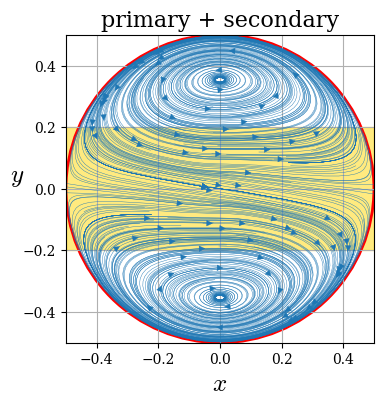}
\caption{\label{fig:1st+2nd} Analytic results for the primary flow without (left) and with (right) the secondary flow.
These results are valid only in the vicinity of the $y = 0$ region (shaded area).}
\end{figure}

Fig.~\ref{fig:1st+2nd} visualizes the analytic result for the primary flow with and without the secondary flow. We set $\Omega=0.03$ and $r_0=0.5$, representative of the slow-orbital-rotation cases shown in Figs.~\ref{fig:4_brcl} and~\ref{fig:phase}. The amplitude $\alpha\rho_0$ remains a free normalization and should be chosen only to give the correct order of magnitude of the velocity field; here we use $\alpha\rho_0=1.0$. The purpose of the figure is therefore not to reproduce the SPH velocity field quantitatively, but to show how the Coriolis term shifts an otherwise symmetric double-cell pattern.

This comparison should be valid  only in  the perturbative regime.  The secondary flows are  evaluated near the $y = 0$ region, and the shaded region in Fig.~\ref{fig:1st+2nd} indicates where this local argument is intended to apply. 
For smaller $J$, however, the circulation is no longer a weakly perturbed symmetric pair. The pressure and density fields are then reorganized by the secondary flow itself, and the linear assumptions used here cease to apply.

\section{Close-Binary Regime and Nonlinear Reorganization}

So far, we have focused on binaries with low orbital angular velocity, corresponding to relatively large  binary separations. In that regime, the numerical results can be interpreted perturbatively in $\Omega$. Our analysis assumed a dipole-like entropy gradient representing a deviation from spherical symmetry. In realistic stars, the temperature distribution is expected to be more  spherical, with lower-order nonsphericities appearing as corrections. The configuration studied above therefore represents an intentionally idealized limit.

Nevertheless,  temperature asymmetry may become more relevant in close binaries as a result of gravity darkening. We therefore turn to the numerical results for close binaries with large orbital angular velocity $\Omega$.

\subsection{Diagnostics of Close-Binary Quasi-Steady States}

\begin{figure}
\includegraphics[width=8.cm]{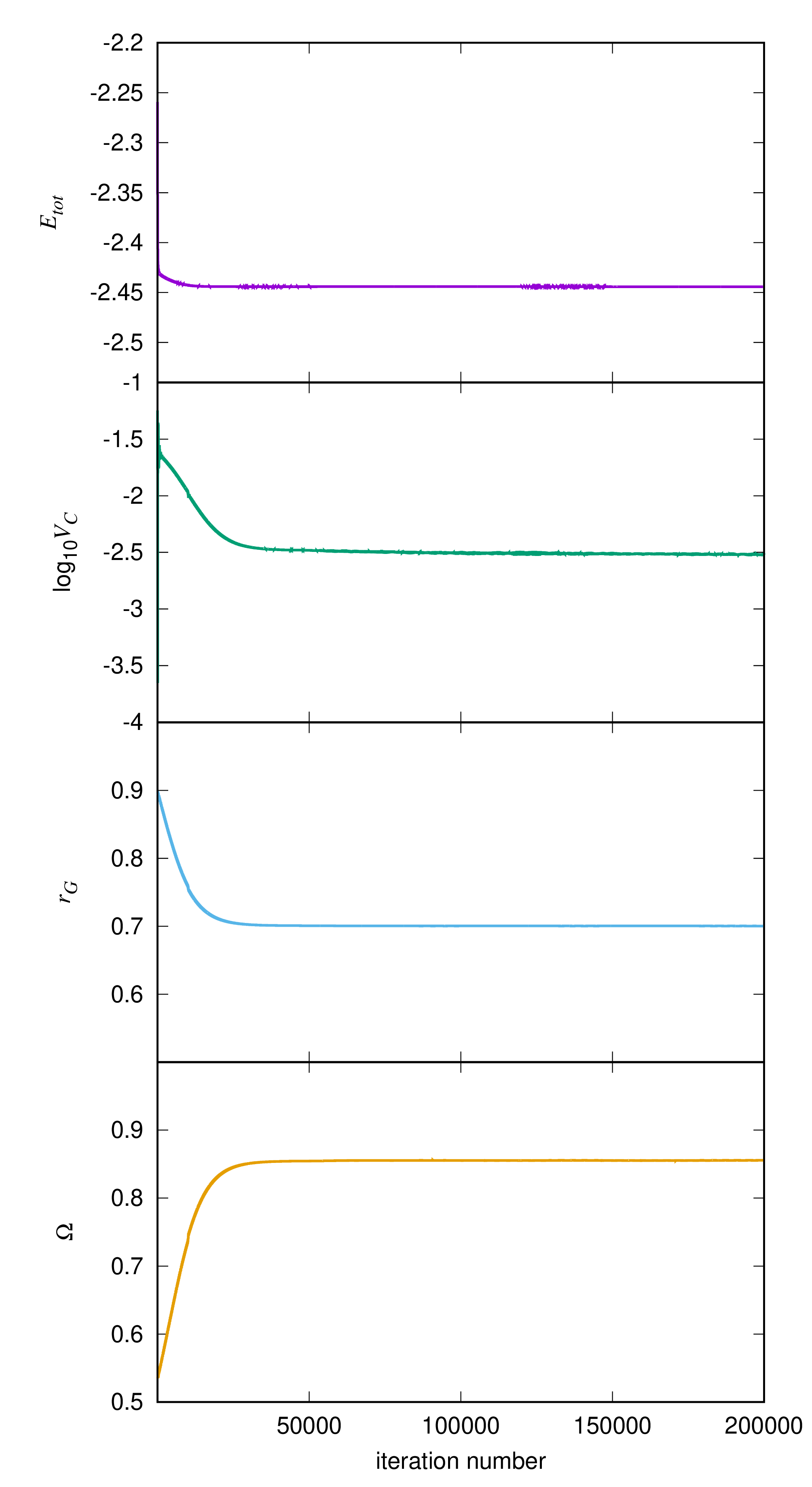}
\caption{\label{fig:n4_brtr} Time evolution of the total energy $E_{\rm tot}$, virial residual $V_C$, orbital distance $r_G$, and orbital angular velocity $\Omega$ for the close-binary barotropic model during relaxation and subsequent free evolution.}
\end{figure}
\begin{figure}
\includegraphics[width=8.cm]{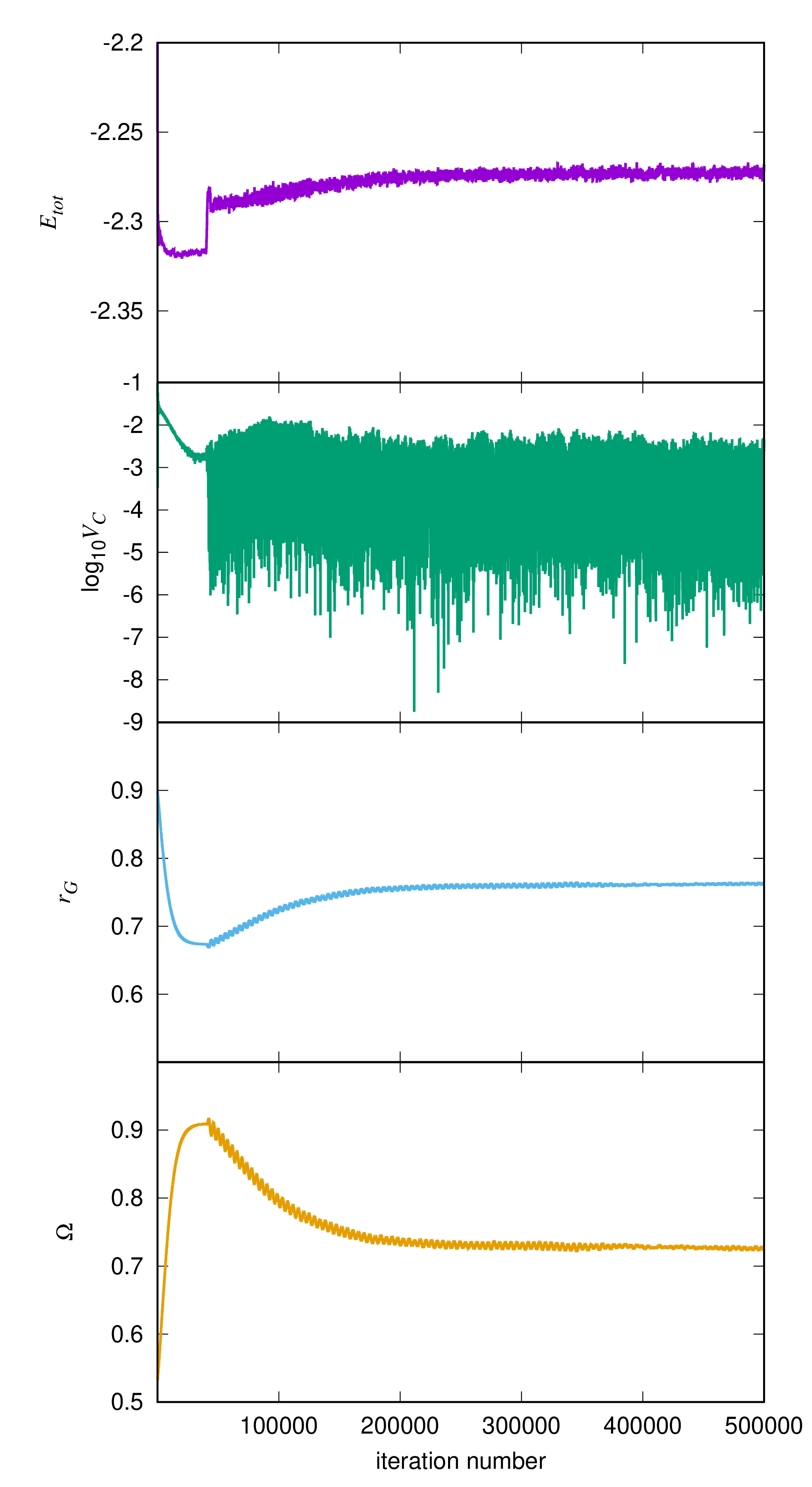}
\caption{\label{fig:n4_brcl}Time evolution of $E_{\rm tot}$, $V_C$, $r_G$, and $\Omega$ for the close-binary baroclinic model during relaxation and subsequent free evolution.}
\end{figure}

As in Section 2, we obtain quasi-steady configurations for both the barotropic and baroclinic cases, but now for close binaries. Both close-binary runs are initialized with nominal target angular momentum $J=0.9$, which is close to  the onset of tidal disruption in the explored model sequence. Toward the end of simulations , the representative values are $r_G\simeq0.70$ and $\Omega\simeq0.86$ in the barotropic run, and $r_G\simeq0.76$ and $\Omega\simeq0.73$ in the baroclinic run. Thus, although the two runs have the same nominal target $J$, their  relaxed structures obtained independently do not have exactly matched orbital parameters, and the comparison below is qualitative. The four diagnostics are again shown in Figs.~\ref{fig:n4_brtr} and~\ref{fig:n4_brcl}. The virial quantity remains at $V_C = O(10^{-3})$, indicating an accuracy comparable to that in Section 3.

A notable feature is that even during the free evolution after 50,000 steps, when the periodic restoration to circular orbits is no longer applied, both cases remain nearly circular. This is evident from the behavior of $r_G$ and $\Omega$, which converge to nearly constant values with only small oscillations. The baroclinic run initially retains a secular drift in $r_G$ after the periodic restoration is stopped. After this drift has subsided, at iteration numbers $\gtrsim 200{,}000$, the remaining radial modulation gives $e_r\lesssim0.01$ according to the estimate given in Eq.~(\ref{eq:e_radial}). 

\subsection{Flow Patterns in Close Binaries}

\begin{figure}
\begin{center}
\includegraphics[width=7.9cm]{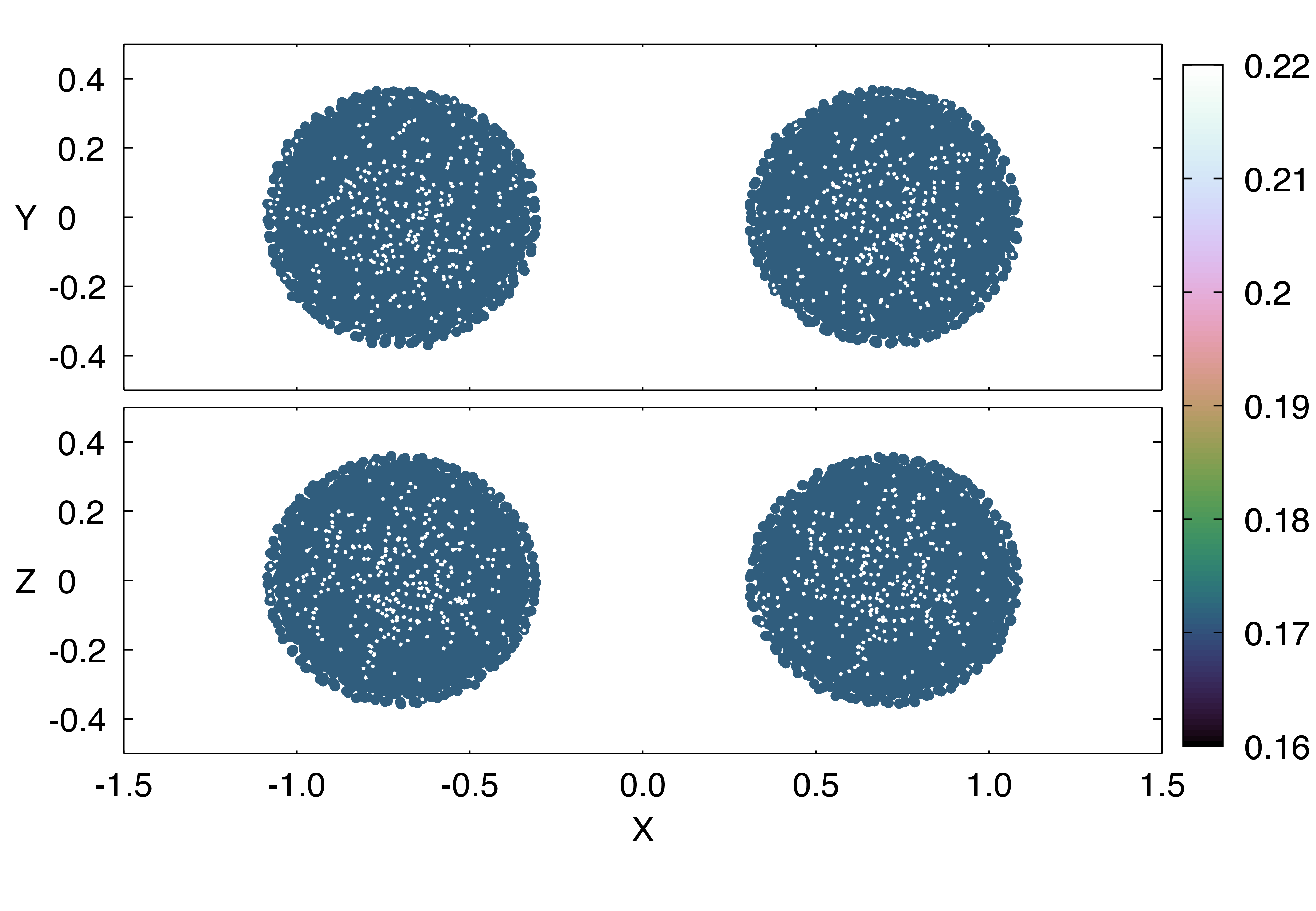}
\caption{\label{fig:nbrtr_K} Resulting $\kappa$ and velocity profiles for the close-binary barotropic case with large $\Omega$. The format and arrow scale are the same as in Fig.~\ref{fig:brtr_K}. }
\end{center}
\end{figure}

\begin{figure}
\begin{center}
\includegraphics[width=7.9cm]{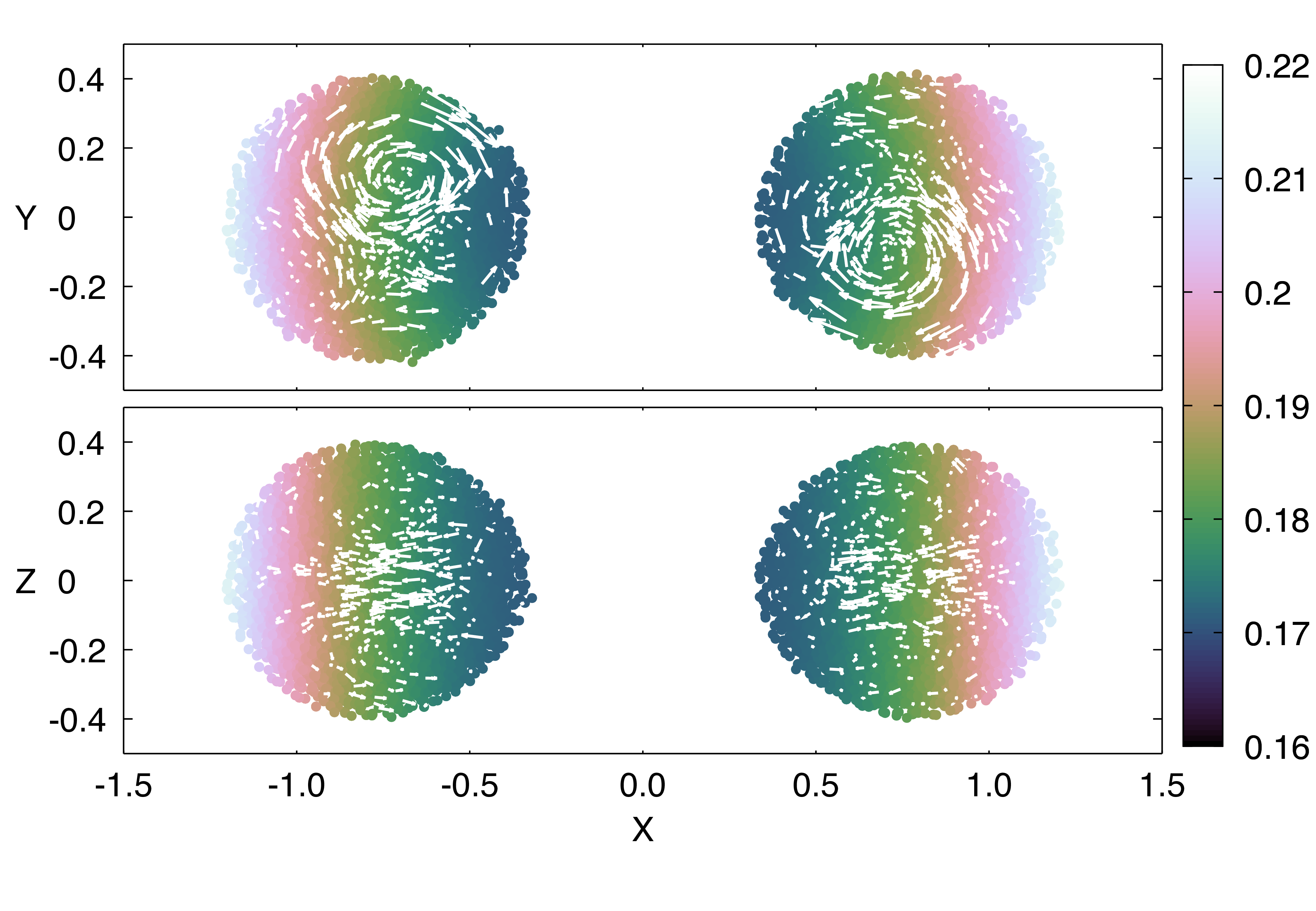}
\caption{\label{fig:nbrcl_K} Resulting $\kappa$ and velocity profiles for the close-binary baroclinic case with large $\Omega$. The format and arrow scale are the same as in Fig.~\ref{fig:nbrtr_K}.}
\end{center}
\end{figure}

In this subsection, we examine the detailed flow structure at the final time steps shown in Figs.~\ref{fig:n4_brtr} and~\ref{fig:n4_brcl}. The snapshots show essentially no orbital-phase dependence, consistent with the nearly circular quasi-steady orbits. Fig.~\ref{fig:nbrtr_K} presents  the barotropic case, while Fig.~\ref{fig:nbrcl_K} shows the baroclinic case.

We first summarize the numerical behavior. As in Fig.~\ref{fig:brtr_K}, the barotropic case exhibits only very small internal velocities compared with the baroclinic case, in which a global flow is driven by the entropy gradient. Because the orbit remains circular, there is essentially no radial velocity associated with the orbital motion, and the residual motion is dominated by random SPH fluctuations. In the baroclinic case, shown in Fig.~\ref{fig:nbrcl_K}, two vortical velocity fields symmetric with respect to $Z = 0$ are formed on the $XZ$-plane, similarly to those seen in Fig.~\ref{fig:brcl_K}. On the $XY$-plane, however, the two vortical velocity fields are asymmetric with respect to $Y = 0$, and the flow is more strongly skewed than in Fig.~\ref{fig:brcl_K}. In this regime, the apparent displacement is no longer straightforward to interpret within the linear picture.

\begin{figure*}
\begin{center}
\includegraphics[width=17.5cm]{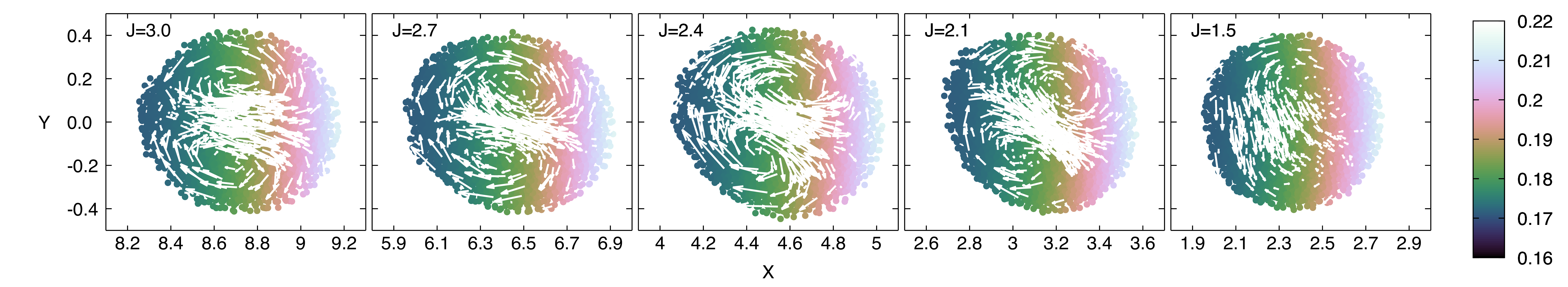}
\caption{\label{fig:Jcompare} Sequence of $XY$-plane slices for independent baroclinic models relaxed with nominal target angular momenta $J=3.0$, $2.7$, $2.4$, $2.1$, and $1.5$ from left to right, focusing on Star-1. The colors show the $\kappa$ profile and the white arrows show the velocity field in the orbital co-rotating frame. Smaller target $J$ corresponds to smaller orbital radius $r_G$ and hence larger orbital angular velocity $\Omega$. As the target $J$ decreases, the skewness of the double-cell circulation grows progressively. In the model with the smallest target angular momentum shown here ($J=1.5$), the flow no longer admits a clear interpretation as a weakly biased pair of symmetric vortices, indicating the breakdown of the perturbative treatment in $\Omega$.}
\end{center}
\end{figure*}

We next examine how the flow pattern depends on the target angular momentum. 
Fig.~\ref{fig:Jcompare} shows a sequence of independent baroclinic models  with the nominal target angular momenta $J=3.0$, $2.7$, $2.4$, $2.1$, and $1.5$ from left to right. A smaller target $J$ corresponds to a smaller orbital radius $r_G$ and hence to a larger orbital angular velocity $\Omega$. The figure summarizes the  flows in the $XY$-plane  in the co-rotating frame  with Star-1 at an origin. As the target $J$ decreases, the  skewness of the double-cell circulation becomes progressively stronger. 
The sequence shows that the orbital-plane skewness strengthens as
the target $J$ decreases.

In the model with the smallest target angular momentum shown here,
$J=1.5$, the perturbative picture breaks down:
$\Omega$ can no longer be regarded as a small perturbation, and the
flow no longer admits a clear interpretation as a weakly biased pair of
symmetric vortices.
 It is clear that the system has entered a nonlinear regime beyond the scope of the  perturbative model presented above. A satisfactory explanation of this behavior will require a treatment beyond the linear regime, including nonlinear hydrodynamical relaxation toward a stable flow.

To quantify the  skewness, we define an asymmetry measure $A$ from the velocity field $(v_x,v_y)$ on the $XY$-plane:
{\small
\begin{eqnarray*}
A &=& \frac{\int \left( |v_x(x,y)-v_x(x,-y)| + |v_y(x,y)+v_y(x,-y)| \right) dS}
         {\int \left( |v_x(x,y)| + |v_x(x,-y)| + |v_y(x,y)| + |v_y(x,-y)| \right) dS}.
\end{eqnarray*}
}
\hspace{-1mm}Here, $dS = dx\,dy$ denotes the area element on the $XY$-plane. Before binning, each snapshot is rotated so that its instantaneous binary axis lies along the $X$-axis, and the rigid orbital velocity estimated from the motion of Star-1 is subtracted. We then select particles with $|z|\leq 0.05$ and evaluate the integrals as discrete sums on a $60 \times 80$ Cartesian grid over a window of width $1.2$ in $x$ and width $0.8$ in $y$, recentered on the current location of Star-1. The velocity in each bin is the unweighted arithmetic mean of the SPH particle velocities in that bin. A pair of bins at $(x,y)$ and $(x,-y)$  contributes to the sum only when both bins are non-empty . All retained pairs have equal area and equal weight, with no density weighting, so the common factor $dS=dx\,dy$ cancels between the numerator and denominator. The quantity $A$ vanishes for perfect reflection symmetry with respect to $Y=0$, and positive values measure the  asymmetry of the double-cell circulation. 
Because the set of paired bins retained in the sums varies among particle numbers and snapshots, finite-particle sampling (shot noise) in the bin occupancy can also contribute to the scatter in $A$.

\begin{figure}
\begin{center}
\includegraphics[width=8.4cm]{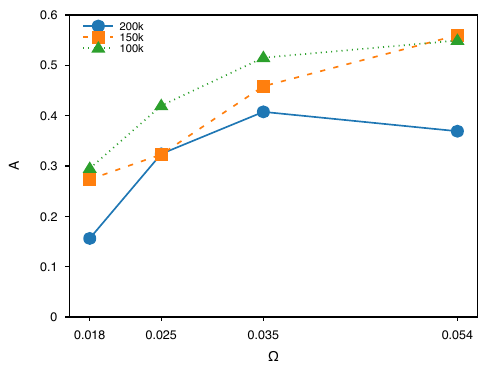}
\caption{\label{fig:Jmetrics} Orbital-plane asymmetry $A$ as a function of the representative orbital angular velocity $\Omega$, evaluated at three total particle numbers in the binary, $N_{\rm tot}=2.0\times 10^5$, $1.5\times 10^5$, and $1.0\times 10^5$. The angular velocities on the horizontal axis, $\Omega=0.018$, $0.025$, $0.035$, and $0.054$, correspond to the target-$J$ models with $J=3.0$, $2.7$, $2.4$, and $2.1$, respectively. Because the compared snapshots have different residual eccentricities and phases, the figure is not used as a strict convergence test.}
\end{center}
\end{figure}

Fig.~\ref{fig:Jmetrics} shows this diagnostic for snapshots selected from the $J=3.0$, $2.7$, $2.4$, and $2.1$ sequences at representative orbital angular velocities $\Omega=0.018$, $0.025$, $0.035$, and $0.054$, respectively. We also compare three total particle numbers, $N_{\rm tot}=2.0\times 10^5$, $1.5\times 10^5$, and $1.0\times 10^5$. This figure should not be interpreted as a precise convergence test: the snapshots come from independently relaxed models and do not have exactly the same residual eccentricity or orbital phase. Consequently, the absolute value of $A$ has limited  significance. The  diagnostic is therefore qualitative. 
In the small-$\Omega$ regime
($\Omega\lesssim0.035$), where the perturbative interpretation is
valid, $A$
 increases monotonically
across the sampled perturbative models.
At larger $\Omega$, 
A tends to be saturated, indicating the onset of the nonlinear regime suggested in Fig.~\ref{fig:Jcompare}. It is also clear, however, that a scatter among the models with different particle numbers gets larger, partly because of the residual eccentricity-dependence; hence the result  should be understood only qualitatively.

\section{Conclusions}
We have studied how imposed baroclinicity and orbital rotation jointly
organize the circulation in the interior of each component in an
equal-mass binary. The goal was not to compute
binary
configurations in mechanical and thermal equilibrium with such internal
flows self-consistently, but to isolate the effect of the orbital rotation on the circulation induced by the (prescribed) thermal asymmetry. We worked in the co-rotating frame of orbital rotation in a controlled way.

Three points emerge from  our model calculations. First,  the 
baroclinicity imposed by the dipolar entropy-gradient develops a
 persistent, well-defined double-cell circulation in  the stellar interiors in the slow-orbital-rotation regime. The analogy with Dean flow is useful but should be interpreted carefully. In curved pipes, the  circulation is generated  in viscous wall-bounded flows and  skewed by the geometric helicity of the conduit. In the binary , by contrast, the circulation is produced by  the entropy gradient imposed by hand in an almost inviscid self-gravitating fluid. 
Moreover, the comparison with Dean flow is limited  to the cross-sectional two-cell topology and its skewness; 
Indeed, the circulation on the
 
orbital plane is different from that on 
 
the meridional plane in the binary

: the circulation remains nearly symmetric with respect to $Z=0$ on the $XZ$-plane, but becomes measurably asymmetric with respect to Y=0 (the direction of the orbital motion) on the $XY$-plane. This  is the  orbital bias we focused on in this paper.

Second, the minimal model we developed in Section 4 reproduces the direction of the bias. It is based on the linear analysis and hence valid only in the slow-orbital-rotation regime. Although it contains a lot of simplifying assumptions, it demonstrated clearly that the Coriolis force induces the bias.

Third, as the orbital angular velocity is increased, the skewness of the double-cell circulation gets progressively stronger on the orbital plane initially.
As the orbital motion becomes even faster, however, the flow pattern starts to deviate from the extrapolation of the trend, indicating the beginning of the nonlinear phase. 
It is also demonstrated by the asymmetry measure $A$ we introduced in Section~5.2: it increases initially
but tends to saturate afterwards at high $\Omega$'s. 
Although the comparison of models with different particle numbers did not show
convergence, those models have different residual eccentricities and
the snapshots used in the comparison have different rotational phases;
the comparison was therefore not meant as a convergence check.

The limitations of the present work should be emphasized. The thermal structure is externally maintained by resetting the entropy distribution repeatedly during the computations. Radiative
 transport and nuclear burning, two important ingredients in 
 thermal balance, are ignored. The resulting configurations  should therefore be interpreted as forced quasi-steady states rather than true  equilibria. In addition, our analysis is restricted to equal-mass binaries and to a particular dipole-like  entropy gradient as an agent of baroclinicity.

From the astrophysical viewpoint, 
the limitations above are closely related to  broader issues in stellar evolution. In the  rotating single star, gravity darkening  is more complex, making the polar regions hotter and the equatorial region cooler in general \citep{EspinosaLara2011}. Rotationally induced transport processes,  such as meridional circulation and shear mixing, must likewise be incorporated  \citep{Ekstrom2012, Georgy2012}, alongside magnetic angular-momentum transport \mbox{\citep{TakahashiLanger2021}} and anisotropic mass loss \mbox{\citep{Hastings2023}}. The present study should  therefore be viewed as a simplified hydrodynamical experiment designed to isolate the flow response to a  thermal asymmetry. An important next step is to extend the same line of investigation to individually rotating stars in binaries, where spin , tidal deformation, and orbital motion should jointly shape the internal circulation.

The present model, albeit overly simplistic, has already identified  a minimal mechanism by which orbital motion  biases the circulation induced by baroclinicity.
  This result should be  a useful starting point for future studies of binary hydrodynamics with more realistic  physics. 

\bigskip

\section*{Acknowledgements}
NY is grateful to R. Igishi, Y. Masada and N. Yokoi for fruitful discussions. SK is grateful to Y. Maekawa for his guidance and valuable discussions. This work was partially supported by JSPS KAKENHI Grant Numbers 24K07054.
The calculations were performed by the supercomputing system HPE SGI 8600 at the Japan Atomic Energy Agency.

\section*{Data availability}
The data underlying this article will be shared on reasonable request to the corresponding author.

\appendix

\section{Coordinate transformation}

\begin{figure}
\begin{center}
\includegraphics[width=8cm]{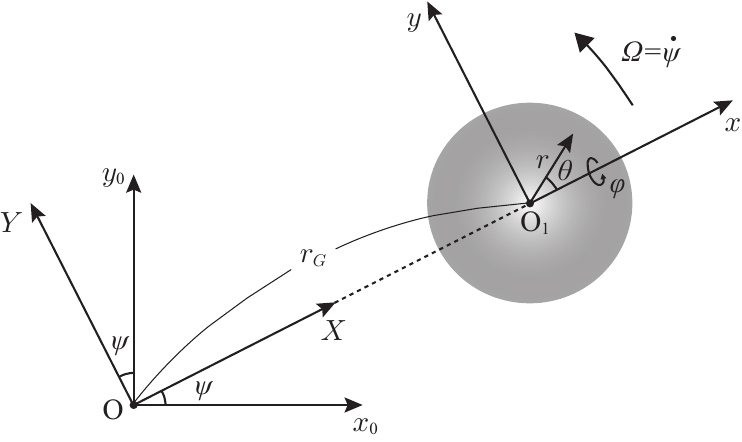}
\caption{\label{fig:coordinate} Coordinate chosen for the linear analysis. The polar axis is the $x$-axis, and $\phi$ denotes the azimuthal angle around the $x$-axis, measured counterclockwise on the $yz$-plane when viewed from the positive $x$ direction.}
\end{center}
\end{figure}

In this appendix, we summarize the coordinate transformation from the inertial frame to the frame co-rotating with the orbital angular velocity. A schematic illustration is given in Fig.~\ref{fig:coordinate}. The numerical calculations are performed in the observer frame, where we adopt a Cartesian coordinate system ${\bf r_0}=(x_0, y_0, z_0)$. In that frame, the fluid obeys the equation of motion
\begin{eqnarray}
\frac{D{\boldsymbol v_0}}{Dt} = -\frac{{\boldsymbol  \nabla} p}{\rho}-{{\boldsymbol  \nabla} \Phi}.
\label{eq:eomA1}
\end{eqnarray}
As in the main text, physical viscous diffusion is neglected.

The two stars orbit around each other, and the center of mass of the binary system is taken to be the origin of the coordinate system. Next, we consider the Lagrangian velocity in the co-rotating frame with orbital angular velocity $\Omega$. The origin of this frame is the same as that of the observer frame.
This coordinate system is identical to the $(X, Y, Z)$ system already introduced in Fig.~\ref{fig:binary_orbit}.
In this case, the following relation holds between the inertial frame and the rotating frame:
\begin{eqnarray}
\left( \frac{D {\boldsymbol R}}{Dt} \right)_I =
\left( \frac{D {\boldsymbol R}}{Dt} \right)_R
+
{\boldsymbol \Omega} \times {\boldsymbol R}.
\end{eqnarray}
Here, the subscripts $I$ and $R$ denote the inertial frame and the rotating frame, respectively, and ${\boldsymbol R}$ is the position vector measured from the origin $O$. More generally, for an arbitrary vector ${\boldsymbol A}$, the following relation holds:
\begin{eqnarray}
\left( \frac{D {\boldsymbol A}}{Dt} \right)_I =
\left( \frac{D {\boldsymbol A}}{Dt} \right)_R
+
{\boldsymbol \Omega} \times {\boldsymbol A}.
\end{eqnarray}

Then we obtain,
\begin{eqnarray}
\left( \frac{D {\boldsymbol v}}{Dt} \right)_I&=&
 \left( \frac{D} {Dt}  \right)_R \left( \frac{D {\boldsymbol R}}{Dt} \right)_I +     {\bf \Omega} \times \left( \frac{D{\boldsymbol R}} {Dt}  \right)_I , \nonumber \\
&=&  \left( \frac{D {\boldsymbol v}}{Dt} \right)_R + 2 {\bf \Omega} \times  \left( \frac{D {\boldsymbol R}}{Dt} \right)_R
 + {\bf \Omega} \times ({\bf \Omega} \times {\boldsymbol  R}). \nonumber \\
\end{eqnarray}
where we have neglected the time derivative of ${\bf \Omega}$.
Therefore, in the rotating frame, Eq.~(\ref{eq:eomA1}) is expressed as 
\begin{eqnarray}
\frac{D{\boldsymbol v}}{Dt} + 2 {\bf \Omega} \times {\boldsymbol  v} + {\bf \Omega} \times ({\bf \Omega} \times {\boldsymbol  R}) = -\frac{{\boldsymbol  \nabla} p}{\rho}-{{\boldsymbol  \nabla} \Phi}, \nonumber \\
\label{eq:eomA5}
\end{eqnarray}
where we have neglected the time derivative of ${\bf \Omega}$.
The effective-potential form in Eq.~(\ref{eq:eom}) follows from Eq.~(\ref{eq:eomA5}) using the following identity:
\begin{eqnarray}
{\bf \Omega} \times ({\bf \Omega} \times {\boldsymbol  R})= -\frac{1}{2} {\boldsymbol  \nabla} \{ ({\bf \Omega} \times {\bf R}) \cdot ({\boldsymbol   \Omega} \times {\boldsymbol   R}) \}.
\end{eqnarray}
Thus, in the binary-centered rotating frame, $\Phi_{\rm eff}=\Phi-\frac{1}{2}({\boldsymbol \Omega}\times{\boldsymbol R})\cdot({\boldsymbol \Omega}\times{\boldsymbol R})$.

Next, we consider shifting the origin to the center of mass of Star-1, denoted by O$_1$. Let ${\boldsymbol r}_1$ be the position vector of O$_1$ measured from O, given by ${\boldsymbol r}_1 = (r_G, 0, 0)$, and assume it to be constant. Then, the position vector can be decomposed as $\boldsymbol{R} = {\boldsymbol r}_1 + {\boldsymbol{r}}$, where ${\boldsymbol{r}}$ is the relative position measured from the center of mass of Star-1, $O_1$. Under this transformation, the equation of motion is obtained by replacing the last term on the left-hand side of Eq.~(\ref{eq:eomA5}). In Eq.~(\ref{eq:eom_star1}), we assume that the gravitational force from Star-2, $\nabla \Phi_2$, is balanced by the centrifugal force ${\bf \Omega} \times ({\bf \Omega} \times {\bf r}_1)$ as
\begin{eqnarray}
 {\bf \Omega} \times ({\bf \Omega} \times {\boldsymbol  r}_1) = -{{\boldsymbol  \nabla} \Phi}_2 . 
\end{eqnarray}

Finally, we introduce spherical coordinates corresponding to the Cartesian coordinates ${\boldsymbol r} = (x, y, z)$ with the origin located at $O_1$. In this transformation, the unit vectors for the spherical $\boldsymbol{\tilde{e}}_i$ is expressed by the unit vectors for the Cartesian coordinates $\boldsymbol{e}_j$ with the rotation transformation matrix ${\boldsymbol U}$ as 
\begin{eqnarray}
\boldsymbol{\tilde{e}}_i = U_{ij} {\boldsymbol e_j}, 
\label{eq:e_sphrcl}
\end{eqnarray}
where the matrix ${\boldsymbol U}$ is given as
\begin{eqnarray}
\boldsymbol{U}= 
\begin{pmatrix}
\sin{\theta} \cos{\phi} & \sin{\theta} \sin{\phi} & \cos{\theta} \\
\cos{\theta} \cos{\phi} & \cos{\theta} \sin{\phi} & -\sin{\theta} \\
-\sin{\phi}                   & \cos{\phi}                  & 0                 \\
\end{pmatrix}.
\end{eqnarray}
Noting that $\theta = 0$ is defined along the $x$-axis, not along the $z$-axis, the indices $i$ and $j$ correspond to $i=r, \theta, \phi$, and $j=y, z, x$, in Eq.~(\ref{eq:e_sphrcl}). 

The Coriolis terms in Eqs.~(\ref{eq:NS_1})--(\ref{eq:NS_3}) are obtained transforming Eq.~(\ref{eq:eom_star1}) via ${\boldsymbol e_i}  = U_{ij}^{-1} \boldsymbol{\tilde{e}}_j$ and $v_i  = U_{ij}^{-1} \tilde{v}_j$ as,

\begin{eqnarray}
\boldsymbol{\Omega} \times \boldsymbol{v}  
&=& -\Omega v_y \boldsymbol{e}_x + \Omega v_x \boldsymbol{e}_y \nonumber \\
&=& -\Omega U_{yl}^{-1}  \tilde{v}_l U_{xm}^{-1} \boldsymbol{\tilde{e}}_m + \Omega U_{xl}^{-1}  \tilde{v}_l U_{ym}^{-1} \boldsymbol{\tilde{e}}_m. 
\end{eqnarray}

The Lagrangian derivative Eq.~(\ref{eq:D}) is also obtained using the analogous relation, $\partial_i  = U_{ij}^{-1} \tilde{\partial}_j$. Hence, the first and second terms in Eqs.~(\ref{eq:NS_1})--(\ref{eq:NS_3}) are immediately derived. Eq.~(\ref{eq:continuity}) is also obtained in the same way.


\label{lastpage}
\end{document}